\documentclass[12pt]{article}
\usepackage{amssymb}
\usepackage{amsmath}
\usepackage{graphicx}
\usepackage{indentfirst}
\usepackage{cite}
\begin{document}

\title{Role of quark-interchange processes in evolution of mesonic matter}
\author{Yu-Qi Li \and Xiao-Ming Xu \and Hui-Jun Ge}
\date{}
\maketitle \vspace{-1cm} {\centerline{\it Department of Physics,
Shanghai University, Baoshan, Shanghai 200444, China}}
\begin{abstract}
We divide the cross section for a meson-meson reaction into three 
parts. The first part is for the quark-interchange process, the second 
for quark-antiquark annihilation processes and the third for resonant
processes. Master rate equations are established to yield time dependence of
fugacities of pions, rhos, kaons and vetor kaons. The equations include cross
sections for inelastic scattering of pions, rhos, kaons and vector kaons.
Cross sections for quark-interchange-induced reactions, that were obtained in
a potential model, are parametrized for convenient use. The number densities of
$\pi$ and $\rho$ ($K$ and $K^\ast$) are altered by quark-interchange processes
in equal magnitudes but opposite signs. The master rate equations 
combined with the hydrodynamic equations for longitudinal and transverse
expansion are solved with many sets of initial meson fugacities.
Quark-interchange processes are shown to be important in the contribution of 
the inelastic meson-meson scattering to evolution of mesonic matter.
\end{abstract}


\newpage
\section{Introduction}
Deconfined matter with high temperature and high density is the focus of 
studies in ultrarelativistic heavy-ion collisions and its confirmation is well 
known to be related to final-state observables. 
But hadronic observables are affected by hadronic
matter that succeeds deconfined matter and 
measurements on dileptons and photons suffer from a background 
that comes from hadronic matter. In order to clearly identify deconfined 
matter from hadronic observables and electromagnetic probes, one has
to subtract any influence of hadronic matter. This forces us to pursue a 
precise description of hadronic matter. In addition, a complete knowledge of 
ultrarelativistic heavy-ion collisions also requires understanding of the 
evolution of hadronic matter.
 
Transport models \cite{RQMD1,RQMD2,ART1,ART2,HSD,UrQMD1,UrQMD2,ARC1,ARC2,
PACIAE1,PACIAE2,Humanic,Nara1,Nara2} can 
provide with vivid and valid descriptions for the evolution of hadronic matter.
The models deal with known and unknown cross
sections for hadron-hadron reactions in the following ways. If experimental 
data are available, parametrizations fitted to the data are first used. If no
measured data exist, cross sections can stem from theoretical calculations, are
simply assumed to be constants, are adapted to the Breit-Wigner formula with
hypothetical widths, or are obtained from the parametrizations
via the detailed balance or via approximate isospin relations. 
Features of a sort of reactions
can be definitely shown by the parametrizations that depend on threshold 
energies and the center-of-mass energy of the two colliding hadrons. The 
assumptions of constant cross sections and the approximate isospin relations
bring uncertainties to study of the evolution of hadronic matter and 
predictions on final-state observables.

Time evolution of meson density, for example, $\phi$ meson density
\cite{Rus02}, can be studied with a rate equation that
includes cross sections for scattering of the meson by other hadrons. The
number of necessary cross sections is huge if rate equations for many hadrons 
are established. To avoid complexity caused by a great number of
mesonic degrees of freedom in hadronic matter, effective numbers of pions and 
kaons were introduced to approximately get evolution of pionic matter and 
kaonic matter \cite{Gav91,Pra99,Son97}. The effective number of pions (kaons) 
is defined as the number of pions (kaons) plus the 
sum of other hadrons weighted by their effective pionic (kaonic) content. For
instance, a $\rho$ meson counts as two pions and an $\Omega$ baryon counts as 
three kaons. With this consideration rate equations for the effective numbers
of pions and kaons were established \cite{Gav91,Pra99,Son97}. 

At mid-rapidity the PHENIX Collaboration \cite{SSAdl04}
obtained ratios of $p_T$-integrated
meson yields in central Au+Au collisions at $\sqrt {s_{NN}}$=200 GeV as 
follows: $\pi^-/\pi^+ = 0.984$, $K^-/K^+ = 0.933$, $K^+/\pi^+ = 0.171$ and 
$K^-/\pi^- = 0.162$, which agree with the results of the other collaborations 
\cite{CAdl04,Ars05,Bac05}. The STAR Collaboration found
$\rho^0/\pi^- =0.169$ in peripheral Au+Au collisions at $\sqrt {s_{NN}}=200$ 
GeV \cite{Ada04}. Obviously, pions, rhos and kaons are dominant meson species
in hadronic matter. In the present work we consider only pions, rhos, kaons
and vector kaons that constitute mesonic matter. Inelastic meson-meson 
scattering originates from resonances, quark-antiquark annihilation processes 
and quark-interchange
processes. At the lowest order a resonance is made up of the remaining quark
and antiquark of which one absorbs a gluon from the annihilation of a quark
in an initial meson and an antiquark in another initial meson. Of course
resonances may be glueballs, multiquark states 
or hybrid states \cite{AT,Bugg}. At the lowest order the
quark-antiquark annihilation means that a quark and an antiquark each in a 
final meson come from the annihilation of a quark and an antiquark each in an
initial meson. A quark-interchange process allows such a type of 
meson-meson scattering where one gluon exchange and
the interchange of two quarks each from an initial meson happen. One is
acquainted with the inelastic meson-meson scattering due to the resonance and 
the quark-antiquark annihilation, but not the scattering induced by the quark 
interchange. One also doesn't know how important the scattering induced by the 
quark interchange is in the evolution of mesonic matter. Indeed, nobody had 
calculated cross sections for the seven quark-interchange-induced reactions 
$\pi \pi \leftrightarrow \rho \rho$ for $I=2$,
$KK \leftrightarrow K^\ast K^\ast$ for $I=1$,
$KK^\ast \leftrightarrow K^\ast K^\ast$ for $I=1$,
$\pi K \leftrightarrow \rho K^\ast$ for $I=\frac {3}{2}$,
$\pi K^\ast \leftrightarrow \rho K^\ast$ for $I=\frac {3}{2}$,
$\rho K \leftrightarrow \rho K^\ast$ for $I=\frac {3}{2}$ and
$\pi K^\ast \leftrightarrow \rho K$ for $I=\frac {3}{2}$, 
until we obtained cross
sections for these reactions in Ref. \cite{Li07} in the quark-interchange
mechanism \cite{Bar92A,Bar92B}. Therefore, we curiously study the role of 
quark-interchange processes that lead to the seven reactions and 
other isospin channels in the evolution of mesonic matter in the present work.
The study resorts to master rate equations for mesons where reactions of pions,
rhos, kaons and vector kaons are taken into account. It will be shown that
quark-interchange processes are important in the contribution of
the inelastic meson-meson scattering to the evolution of mesonic matter. 
Therefore, if we include the resonances and the quark-antiquark
annihilation processes, the quark-interchange processes should be included on 
an equal footing.
 
In the next section the master rate equations 
for pions, rhos, kaons and vector kaons are established while the inelastic 
meson-meson scattering due to the resonance, the quark-antiquark
annihilation and the quark interchange is considered. In Section 3
parametrizations of cross sections for the quark-interchange-induced reactions
are presented. Cross sections for the quark-antiquark annihilation processes
and the resonant processes are individually introduced. Numerical results of 
the master rate equations associated with longitudinal expansion and 
discussions are given in Section 4. The master rate equations are extended to
include $2 \leftrightarrow 1$ mesonic
processes in Section 5 and in the case of both
the longitudinal and transverse expansion the importance of the 
quark-interchange processes is examined in Section 6. Summary is in the last 
section.

\section{Master rate equations}
We establish the notation 
$K=\left( \begin{array}{c}K^+\\ K^0 \end{array} \right)$ 
and $\bar{K}=\left( \begin{array}{c}\bar{K}^0\\ K^- \end{array} \right)$ for
the pseudoscalar isospin doublets as well as
$K^{\ast}=\left( \begin{array}{c}K^{\ast +}\\ K^{\ast 0} \end{array} \right)$ 
and $\bar{K}^{\ast}=\left( \begin{array}{c}\bar{K}^{\ast 0}\\ K^{\ast -} 
\end{array} \right)$ for the vector isospin doublets.
We only consider $\pi$, $\rho$, $K$, $\bar K$, $K^\ast$ and $\bar {K}^\ast$ of
mesonic matter. In order to clearly exhibit the role of the 
quark-interchange processes, we neglect both $2 \to 1$ mesonic reactions
and decays of $\rho$, $K^\ast$ and $\bar {K}^\ast$ in this section and the 
next two sections. The negligence does not affect us to draw correct 
conclusions. The inclusion of the $2 \leftrightarrow 1$ mesonic processes is 
deferred to Section 5. The reactions that can change the numbers of
$\pi$, $\rho$, $K$, $\bar K$, $K^\ast$ and $\bar {K}^\ast$ in a lifetime of
mesonic matter are the following inelastic 2-to-2 scattering:
\begin{enumerate}
\item $\pi \pi \leftrightarrow \rho \rho$,
\item $K K \leftrightarrow K^\ast K^\ast$ and $\bar{K} \bar{K} \leftrightarrow 
\bar{K}^\ast \bar{K}^\ast$,
\item $K K^\ast \leftrightarrow K^\ast K^\ast$ and $\bar{K} \bar{K}^\ast 
\leftrightarrow \bar{K}^\ast \bar{K}^\ast$,
\item $K \bar{K} \leftrightarrow K^\ast \bar{K}^\ast$,
\item $K \bar{K}^\ast \leftrightarrow K^\ast \bar{K}^\ast$ and $K^\ast \bar{K} 
\leftrightarrow K^\ast \bar{K}^\ast$,
\item $\pi K^\ast \leftrightarrow \rho K$ and $\pi \bar{K}^\ast 
\leftrightarrow \rho \bar{K}$,
\item $\pi K \leftrightarrow \rho K^\ast$ and $\pi \bar{K} \leftrightarrow 
\rho \bar{K}^\ast$,
\item $\pi K^\ast \leftrightarrow \rho K^\ast$ and $\pi \bar{K}^\ast 
\leftrightarrow \rho \bar{K}^\ast$,
\item $\rho K \leftrightarrow \rho K^\ast$ and $\rho \bar{K} \leftrightarrow 
\rho \bar{K}^\ast$,
\item $\pi \pi \leftrightarrow K \bar{K}$,
\item $\pi \rho \leftrightarrow K \bar{K}^\ast$ and $\pi \rho \leftrightarrow 
K^\ast \bar{K}$,
\item $K \bar{K} \leftrightarrow \rho \rho$.
\end{enumerate}
The cross sections for these reactions are not independent of each other, 
e.g., 
$\sigma_{\bar{K} \bar{K} \rightarrow \bar{K}^\ast \bar{K}^\ast}
=\sigma_{K K\rightarrow K^\ast K^\ast}$, 
$\sigma_{\bar{K} \bar{K}^\ast \rightarrow \bar{K}^\ast \bar{K}^\ast}
=\sigma_{K K^\ast \rightarrow K^\ast K^\ast}$, 
$\sigma_{K^\ast \bar{K} \rightarrow K^\ast \bar{K}^\ast}
=\sigma_{K \bar{K}^\ast \rightarrow K^\ast \bar{K}^\ast}$, 
$\sigma_{\pi \bar{K}^\ast \rightarrow \rho \bar{K}}
=\sigma_{\pi K^\ast \rightarrow \rho K}$, 
$\sigma_{\pi \bar{K} \rightarrow \rho \bar{K}^\ast}
=\sigma_{\pi K \rightarrow \rho K^\ast}$, 
$\sigma_{\pi \bar{K}^\ast \rightarrow \rho \bar{K}^\ast}
=\sigma_{\pi K^\ast \rightarrow \rho K^\ast}$,
$\sigma_{\rho \bar{K} \rightarrow \rho \bar{K}^\ast}
=\sigma_{\rho K \rightarrow \rho K^\ast}$, 
and 
$\sigma_{\pi \rho \rightarrow K^\ast \bar{K}}
=\sigma_{\pi \rho \rightarrow K \bar{K}^\ast}$.

Meson number densities change with time according to the following
rate equations,
\begin{equation} \label{rate}
\partial_{\mu}(n_{i}u^{\mu})=\Psi_{i},
\end{equation}
where $u^{\mu}=(u^0,\vec {u})=\gamma(1,\vec{\rm v})$ is the four-velocity of 
the local reference frame comoving at velocity $\vec{\rm v}$ and with the 
Lorentz 
factor $\gamma$. $n_\pi$, $n_\rho$, $n_K$, $n_{\bar K}$, $n_{K^\ast}$ and 
$n_{\bar {K}^\ast}$ are the number densities of $\pi$, $\rho$, $K$, $\bar K$, 
$K^\ast$ and $\bar {K}^\ast$ if $i$ denotes $\pi$, $\rho$, $K$, $\bar K$, 
$K^\ast$ and $\bar {K}^\ast$, respectively.
Zero values of the source terms $\Psi_{i}$ 
mean that the total number of each particle species is conserved.
The source terms are given by

\begin{align}\label{eqpi}
\Psi_{\pi}=&2\times\frac{1}{2}\langle\sigma_{\rho\rho\rightarrow\pi\pi}
v_{\rho\rho}\rangle n_{\rho}^2
-2\times\frac{1}{2}\langle\sigma_{\pi\pi\rightarrow\rho\rho}v_{\pi\pi}\rangle 
n_{\pi}^2 \notag \\
&+\langle\sigma_{\rho K\rightarrow\pi K^{\ast}}v_{\rho K}\rangle n_{\rho}n_{K}
-\langle\sigma_{\pi K^{\ast}\rightarrow\rho K}v_{\pi K^{\ast}}\rangle n_{\pi}
n_{K^{\ast}} \notag \\
&+\langle\sigma_{\rho \bar{K}\rightarrow\pi \bar{K}^{\ast}}v_{\rho \bar{K}}
\rangle n_{\rho}n_{\bar{K}}
-\langle\sigma_{\pi \bar{K}^{\ast}\rightarrow\rho \bar{K}}
v_{\pi \bar{K}^{\ast}}
  \rangle n_{\pi}n_{\bar{K}^{\ast}} \notag \\
&+\langle\sigma_{\rho K^{\ast}\rightarrow\pi K}v_{\rho {K}^\ast}\rangle 
n_{\rho}n_{K^{\ast}}
-\langle\sigma_{\pi K\rightarrow\rho K^{\ast}}v_{\pi K}\rangle n_{\pi}n_{K} 
\notag \\
&+\langle\sigma_{\rho \bar{K}^{\ast}\rightarrow\pi \bar{K}}
v_{\rho \bar{K}^\ast}\rangle n_{\rho}n_{\bar{K}^{\ast}}
-\langle\sigma_{\pi \bar{K}\rightarrow\rho \bar{K}^{\ast}}v_{\pi \bar{K}}
\rangle n_{\pi}n_{\bar{K}} \notag \\
&+\langle\sigma_{\rho K^{\ast}\rightarrow\pi K^{\ast}}v_{\rho K^{\ast}}
\rangle n_{\rho}n_{K^{\ast}}
-\langle\sigma_{\pi K^{\ast}\rightarrow\rho K^{\ast}}v_{\pi K^{\ast}}
\rangle n_{\pi}n_{K^{\ast}} \notag \\
&+\langle\sigma_{\rho \bar{K}^{\ast}\rightarrow\pi \bar{K}^{\ast}}v_{\rho 
\bar{K}^{\ast}}\rangle
  n_{\rho}n_{\bar{K}^{\ast}}
-\langle\sigma_{\pi \bar{K}^{\ast}\rightarrow\rho \bar{K}^{\ast}}v_{\pi 
\bar{K}^{\ast}}\rangle
  n_{\pi}n_{\bar{K}^{\ast}} \notag \\
&+2\langle\sigma_{K\bar{K}\rightarrow\pi\pi}v_{K\bar{K}}\rangle n_{K}
n_{\bar{K}}
-2\times\frac{1}{2}\langle\sigma_{\pi\pi\rightarrow K\bar{K}}v_{\pi\pi}\rangle 
n_{\pi}^2 \notag \\
&+\langle\sigma_{K\bar{K}^{\ast}\rightarrow\pi \rho}v_{K\bar{K}^{\ast}}\rangle 
n_{K}n_{\bar{K}^{\ast}}
-\langle\sigma_{\pi \rho\rightarrow K\bar{K}^{\ast}}v_{\pi \rho}\rangle n_{\pi}
n_{\rho} \notag \\
&+\langle\sigma_{K^{\ast}\bar{K}\rightarrow\pi \rho}v_{K^{\ast}\bar{K}}\rangle 
n_{K^{\ast}}n_{\bar{K}}
-\langle\sigma_{\pi \rho\rightarrow K^{\ast}\bar{K}}v_{\pi \rho}\rangle n_{\pi}
n_{\rho},
\end{align}
\begin{align}\label{eqrho}
\Psi_{\rho}=&2\times\frac{1}{2}\langle\sigma_{\pi\pi\rightarrow\rho\rho}
v_{\pi\pi}\rangle n_{\pi}^2
-2\times\frac{1}{2}\langle\sigma_{\rho\rho\rightarrow\pi\pi}v_{\rho\rho}
\rangle n_{\rho}^2 \notag \\
&+\langle\sigma_{\pi K^{\ast}\rightarrow\rho K}v_{\pi K^{\ast}}\rangle n_{\pi}
n_{K^{\ast}}
-\langle\sigma_{\rho K\rightarrow\pi K^{\ast}}v_{\rho K}\rangle n_{\rho}n_{K} 
\notag \\
&+\langle\sigma_{\pi \bar{K}^{\ast}\rightarrow\rho \bar{K}}
v_{\pi \bar{K}^{\ast}}\rangle n_{\pi}n_{\bar{K}^{\ast}}
-\langle\sigma_{\rho \bar{K}\rightarrow\pi \bar{K}^{\ast}}
v_{\rho \bar{K}}\rangle n_{\rho}n_{\bar{K}} \notag \\
&+\langle\sigma_{\pi K\rightarrow\rho K^{\ast}}v_{\pi K}\rangle n_{\pi}n_{K}
-\langle\sigma_{\rho K^{\ast}\rightarrow\pi K}v_{\rho K^\ast}\rangle n_{\rho}
n_{K^{\ast}} \notag \\
&+\langle\sigma_{\pi \bar{K}\rightarrow\rho \bar{K}^{\ast}}
v_{\pi \bar{K}}\rangle n_{\pi}n_{\bar{K}}
-\langle\sigma_{\rho \bar{K}^{\ast}\rightarrow\pi \bar{K}}
v_{\rho \bar{K}^\ast}\rangle n_{\rho}n_{\bar{K}^{\ast}} \notag \\
&+\langle\sigma_{\pi K^{\ast}\rightarrow\rho K^{\ast}}
v_{\pi K^{\ast}}\rangle n_{\pi}n_{K^{\ast}}
-\langle\sigma_{\rho K^{\ast}\rightarrow\pi K^{\ast}}
v_{\rho K^{\ast}}\rangle n_{\rho}n_{K^{\ast}} \notag \\
&+\langle\sigma_{\pi \bar{K}^{\ast}\rightarrow\rho \bar{K}^{\ast}}
v_{\pi \bar{K}^{\ast}}\rangle
  n_{\pi}n_{\bar{K}^{\ast}}
-\langle\sigma_{\rho \bar{K}^{\ast}\rightarrow\pi \bar{K}^{\ast}}
v_{\rho \bar{K}^{\ast}}\rangle
  n_{\rho}n_{\bar{K}^{\ast}} \notag \\
&+\langle\sigma_{K\bar{K}^{\ast}\rightarrow\pi \rho}v_{K\bar{K}^{\ast}}\rangle 
n_{K}n_{\bar{K}^{\ast}}
-\langle\sigma_{\pi \rho\rightarrow K\bar{K}^{\ast}}v_{\pi \rho}\rangle n_{\pi}
n_{\rho} \notag \\
&+\langle\sigma_{K^{\ast}\bar{K}\rightarrow\pi \rho}v_{K^{\ast}\bar{K}}\rangle 
n_{K^{\ast}}n_{\bar{K}}
-\langle\sigma_{\pi \rho\rightarrow K^{\ast}\bar{K}}v_{\pi \rho}\rangle n_{\pi}
n_{\rho} \notag \\
&+2\langle\sigma_{K\bar{K}\rightarrow\rho\rho}v_{K\bar{K}}\rangle n_{K}
n_{\bar{K}}
-2\times\frac{1}{2}\langle\sigma_{\rho\rho\rightarrow K\bar{K}}v_{\rho\rho}
\rangle n_{\rho}^2,
\end{align}
\begin{align}\label{eqk}
\Psi_{K}=&2\times\frac{1}{2}\langle\sigma_{K^{\ast} K^{\ast}\rightarrow KK}
v_{K^{\ast} K^{\ast}}\rangle n_{K^{\ast}}^2
-2\times\frac{1}{2}\langle\sigma_{KK \rightarrow K^{\ast} K^{\ast}}v_{KK}
\rangle n_{K}^2 \notag \\
&+\frac{1}{2}\langle\sigma_{K^{\ast} K^{\ast}\rightarrow KK^{\ast}}v_{K^{\ast} 
K^{\ast}}\rangle n_{K^{\ast}}^2
-\langle\sigma_{KK^{\ast} \rightarrow K^{\ast} K^{\ast}}v_{KK^{\ast}}\rangle 
n_{K}n_{K^{\ast}} \notag \\
&+\langle\sigma_{K^{\ast}\bar{K}^{\ast}\rightarrow K\bar{K}}v_{K^{\ast} 
\bar{K}^{\ast}}\rangle
  n_{K^{\ast}}n_{\bar{K}^{\ast}}
-\langle\sigma_{K\bar{K} \rightarrow K^{\ast}\bar{K}^{\ast}}v_{K\bar{K}}
\rangle n_{K}n_{\bar{K}} \notag \\
&+\langle\sigma_{K^{\ast}\bar{K}^{\ast}\rightarrow K\bar{K}^{\ast}}
v_{K^{\ast}\bar{K}^{\ast}}\rangle
  n_{K^{\ast}}n_{\bar{K}^{\ast}}
-\langle\sigma_{K\bar{K}^{\ast}\rightarrow K^{\ast}\bar{K}^{\ast}}
v_{K\bar{K}^{\ast}}\rangle
  n_{K}n_{\bar{K}^{\ast}} \notag \\
&+\langle\sigma_{\pi K^{\ast}\rightarrow\rho K}v_{\pi K^{\ast}}\rangle 
n_{\pi}n_{K^{\ast}}
-\langle\sigma_{\rho K\rightarrow\pi K^{\ast}}v_{\rho K}\rangle n_{\rho}n_{K} 
\notag \\
&+\langle\sigma_{\rho K^{\ast}\rightarrow\pi K}v_{\rho K^\ast}\rangle n_{\rho}
n_{K^{\ast}}
-\langle\sigma_{\pi K\rightarrow\rho K^{\ast}}v_{\pi K}\rangle n_{\pi}n_{K} 
\notag \\
&+\langle\sigma_{\rho K^{\ast}\rightarrow\rho K}v_{\rho K^\ast}\rangle n_{\rho}
n_{K^{\ast}}
-\langle\sigma_{\rho K\rightarrow\rho K^{\ast}}v_{\rho K}\rangle n_{\rho}n_{K} 
\notag \\
&+\frac{1}{2}\langle\sigma_{\pi\pi\rightarrow K\bar{K}}v_{\pi\pi}\rangle 
n_{\pi}^2
-\langle\sigma_{K\bar{K}\rightarrow\pi\pi}v_{K\bar{K}}\rangle n_{K}n_{\bar{K}} 
\notag \\
&+\langle\sigma_{\pi \rho\rightarrow K\bar{K}^{\ast}}v_{\pi \rho}\rangle 
n_{\pi}n_{\rho}
-\langle\sigma_{K\bar{K}^{\ast}\rightarrow\pi \rho}v_{K\bar{K}^{\ast}}\rangle 
n_{K}n_{\bar{K}^{\ast}} \notag \\
&+\frac{1}{2}\langle\sigma_{\rho\rho\rightarrow K\bar{K}}v_{\rho\rho}\rangle 
n_{\rho}^2
-\langle\sigma_{K\bar{K}\rightarrow\rho\rho}v_{K\bar{K}}\rangle n_{K}
n_{\bar{K}},
\end{align}
\begin{align}\label{eqkstar}
\Psi_{K^\ast}=&2\times\frac{1}{2}\langle\sigma_{KK \rightarrow K^{\ast} 
K^{\ast}}v_{KK}\rangle n_{K}^2
-2\times\frac{1}{2}\langle\sigma_{K^{\ast} K^{\ast}\rightarrow KK}
v_{K^{\ast} K^{\ast}}\rangle n_{K^{\ast}}^2 \notag \\
&+\langle\sigma_{KK^{\ast} \rightarrow K^{\ast} K^{\ast}}v_{KK^{\ast}}\rangle 
n_{K}n_{K^{\ast}}
-\frac{1}{2}\langle\sigma_{K^{\ast} K^{\ast}\rightarrow KK^{\ast}}v_{K^{\ast} 
K^{\ast}}\rangle n_{K^{\ast}}^2 \notag \\
&+\langle\sigma_{K\bar{K} \rightarrow K^{\ast}\bar{K}^{\ast}}v_{K\bar{K}}
\rangle n_{K}n_{\bar{K}}
-\langle\sigma_{K^{\ast}\bar{K}^{\ast}\rightarrow K\bar{K}}v_{K^{\ast} 
\bar{K}^{\ast}}\rangle
  n_{K^{\ast}}n_{\bar{K}^{\ast}} \notag \\
&+\langle\sigma_{K\bar{K}^{\ast}\rightarrow K^{\ast}\bar{K}^{\ast}}
v_{K\bar{K}^{\ast}}\rangle
  n_{K}n_{\bar{K}^{\ast}}
-\langle\sigma_{K^{\ast}\bar{K}^{\ast}\rightarrow K\bar{K}^{\ast}}
v_{K^{\ast}\bar{K}^{\ast}}\rangle
  n_{K^{\ast}}n_{\bar{K}^{\ast}} \notag \\
&+\langle\sigma_{\rho K\rightarrow\pi K^{\ast}}v_{\rho K}\rangle n_{\rho}n_{K}
-\langle\sigma_{\pi K^{\ast}\rightarrow\rho K}v_{\pi K^{\ast}}\rangle n_{\pi}
n_{K^{\ast}} \notag \\
&+\langle\sigma_{\pi K\rightarrow\rho K^{\ast}}v_{\pi K}\rangle n_{\pi}n_{K}
-\langle\sigma_{\rho K^{\ast}\rightarrow\pi K}v_{\rho K^\ast}\rangle n_{\rho}
n_{K^{\ast}} \notag \\
&+\langle\sigma_{\rho K\rightarrow\rho K^{\ast}}v_{\rho K}\rangle n_{\rho}n_{K}
-\langle\sigma_{\rho K^{\ast}\rightarrow\rho K}v_{\rho K^\ast}\rangle n_{\rho}
n_{K^{\ast}} \notag \\
&+\langle\sigma_{\pi \rho\rightarrow K^{\ast}\bar{K}}v_{\pi \rho}\rangle 
n_{\pi}n_{\rho}
-\langle\sigma_{K^{\ast}\bar{K}\rightarrow\pi \rho}v_{K^{\ast}\bar{K}}\rangle 
n_{K^{\ast}}n_{\bar{K}}.
\end{align}
The source term of $\bar K$ ($\bar {K}^\ast$) is not shown since 
$\Psi_{\bar K}$ ($\Psi_{\bar {K}^*}$) is obtained from $\Psi_K$ ($\Psi_{K^*}$) 
by the replacements of the subscripts, $K \leftrightarrow \bar K$ and $K^* 
\leftrightarrow \bar{K}^\ast$. 
Those terms what contain the factor 2 relate to the reactions that have two 
indistinguishable initial or final mesons. This factor means that the 
two initial mesons of a species vanish to attain no final mesons of the species
or the two final mesons of a species appear from no initial mesons of the 
species. The first fourteen terms of $\Psi_\pi$ equal the negative of
the first fourteen terms of $\Psi_\rho$. The first fourteen terms of $\Psi_K$
equal the negative of the first fourteen terms of $\Psi_{K^\ast}$. The
quark-interchange processes are only contained in these terms. Therefore, the 
quark-interchange processes contribute to the variations of 
the number densities of
$\pi$ and $\rho$, or $K$ and $K^\ast$, in equal magnitudes but opposite signs.

The thermal averaged cross section with
the relative velocity of two initial mesons $v_{\rm rel}$ is defined as
\begin{equation}\label{avercross}
\langle \sigma_{ij\to i^\prime j^\prime} v_{\rm{rel}}\rangle
=\frac{\int\frac{d^3 k_1}{(2\pi)^3}f_i(k_1)
\frac{d^3 k_2}{(2\pi)^3}f_j(k_2)\sigma_{ij \to i^\prime j^\prime}
(\sqrt{s})v_{\rm{rel}}}
{\int\frac{d^3 k_1}{(2\pi)^3}f_i(k_1)\int\frac{d^3 k_2}{(2\pi)^3}
f_j(k_2)},
\end{equation}
where $f_i(k_1)$ and $f_j(k_2)$ are the momentum distributions of the
two initial mesons with the four-momenta $k_1$ and $k_2$,
respectively; $\sigma_{ij \to i^\prime j^\prime}(\sqrt{s})$ is a
cross section that depends on the center-of-mass energy $\sqrt s$
of the two initial mesons.
We take the approximate factorization form of the J\"{u}ttner distribution 
with nonequilibrium fugacity $\lambda_i$ of particle species $i$,
\begin{equation}\label{distribution}
f_i(k) =\frac{\lambda_i}{e^{u\cdot k/T}-1},
\end{equation}
where $T$ is temperature. If two initial mesons are indistinguishable, $f_i$
and $f_j$ possess the same fugacity and as seen in Eqs. (2)-(5) a factor of 
$\frac {1}{2}$ is in some terms to remove the double counting of initial mesons
in the thermal average.

The number density of particle species $i$ is given by
\begin{equation}\label{ndensity1}
n_i=g_i\int\frac{d^3 k}{(2\pi)^3}\frac{\lambda_i}{e^{u\cdot
k/T}-1}=u^0 \lambda_i \bar{n}_i,
\end{equation}
with
\begin{equation}\label{ndensity2}
\bar{n}_i=\frac{g_i}{2\pi^2}\int^\infty_0 d
|\vec{k}^{'}|\frac{\vec{k}^{'2}}{e^{\sqrt{\vec{k}^{'2}+m^2_i}/T}-1},
\end{equation}
where $m_i$ is the mass of particle species $i$; the spin-isospin degeneracy 
factor $g_i=3$ for $\pi$, 9 for $\rho$, 2 for $K$ or $\bar K$, 6 for $K^\ast$
or $\bar {K}^\ast$; $\vec {k}^\prime$ is the
particle momentum in the local comoving reference frame.
The derivative of $\bar{n}_i$ with respect to $T$ is
\begin{equation}\label{}
\frac{d \bar{n}_i}{d T}=\frac{1}{T}\left(3 \bar{n}_i+\bar{n}_{i-}
\right),
\end{equation}
with
\begin{equation}\label{}
\bar{n}_{i-}=\frac{g_i}{2\pi^2}\int^\infty_0 d
|\vec{k}^{'}|\frac{m^2_i}{e^{\sqrt{\vec{k}^{'2}+m^2_i}/T}-1}.
\end{equation}
For symmetric matter $\lambda_{\bar K}=\lambda_K$, $\lambda_{\bar {K}^\ast}=
\lambda_{K^\ast}$, $\bar {n}_{\bar K}=\bar {n}_K$, $\bar {n}_{\bar {K}^\ast}
=\bar {n}_{K^\ast}$, $\bar {n}_{\bar {K}-}=\bar {n}_{K-}$ and
$\bar {n}_{\bar {K}^\ast -}=\bar {n}_{K^\ast -}$. 

Inserting Eqs. (8) and (10) into Eqs. (2)-(5), we obtain rate equations for
fugacities of $\pi$, $\rho$, $K$ and $K^\ast$ of symmetric matter
in the longitudinal expansion
\begin{align}\label{eqpi2}
&\frac{\dot{\lambda}_\pi}{\lambda_\pi}+\left(3+\frac{\bar{n}_{\pi-}}
{\bar{n}_\pi}\right)\frac{\dot{T}}{T}
 +\frac{1}{\tau} \notag \\
&=2\times\frac{1}{2}\langle\sigma_{\rho\rho\rightarrow\pi\pi}
v_{\rho\rho}\rangle
  u^0\frac{\lambda^2_\rho \bar{n}^2_\rho}{\lambda_\pi \bar{n}_\pi}
 -2\times\frac{1}{2}\langle\sigma_{\pi\pi\rightarrow\rho\rho}v_{\pi\pi}
\rangle u^0\lambda_\pi\bar{n}_\pi \notag \\
&~~~+2\langle\sigma_{\rho K\rightarrow\pi K^{\ast}}v_{\rho K}\rangle
  u^0\frac{\lambda_\rho \bar{n}_\rho\lambda_K \bar{n}_K}{\lambda_\pi 
\bar{n}_\pi}
 -2\langle\sigma_{\pi K^{\ast}\rightarrow\rho K}v_{\pi K^{\ast}}\rangle 
u^0\lambda_{K^\ast} \bar{n}_{K^\ast} \notag \\
&~~~+2\langle\sigma_{\rho K^{\ast}\rightarrow\pi K}v_{\rho K^\ast}\rangle
  u^0\frac{\lambda_\rho \bar{n}_\rho\lambda_{K^\ast} \bar{n}_{K^\ast}}
{\lambda_\pi \bar{n}_\pi}
 -2\langle\sigma_{\pi K\rightarrow\rho K^{\ast}}v_{\pi K}\rangle u^0\lambda_K 
\bar{n}_K \notag \\
&~~~+2\langle\sigma_{\rho K^{\ast}\rightarrow\pi K^{\ast}}
v_{\rho K^{\ast}}\rangle
  u^0\frac{\lambda_\rho \bar{n}_\rho\lambda_{K^\ast} \bar{n}_{K^\ast}}
{\lambda_\pi \bar{n}_\pi}
 -2\langle\sigma_{\pi K^{\ast}\rightarrow\rho K^{\ast}}v_{\pi K^{\ast}}\rangle 
u^0\lambda_{K^\ast}\bar{n}_{K^\ast}\notag\\
&~~~+2\langle\sigma_{K\bar{K}\rightarrow\pi\pi}v_{K\bar{K}}\rangle
  u^0\frac{\lambda^2_K \bar{n}^2_K}{\lambda_\pi \bar{n}_\pi}
 -2\times\frac{1}{2}\langle\sigma_{\pi\pi\rightarrow K\bar{K}}
v_{\pi\pi}\rangle u^0\lambda_\pi\bar{n}_\pi \notag \\
&~~~+2\langle\sigma_{K\bar{K}^{\ast}\rightarrow\pi \rho}
v_{K\bar{K}^{\ast}}\rangle
  u^0\frac{\lambda_K \bar{n}_K\lambda_{K^\ast} \bar{n}_{K^\ast}}
{\lambda_\pi \bar{n}_\pi}
 -2\langle\sigma_{\pi \rho\rightarrow K\bar{K}^{\ast}}
v_{\pi \rho}\rangle u^0\lambda_\rho \bar{n}_\rho,
\end{align}
\begin{align}\label{eqrho2}
&\frac{\dot{\lambda}_\rho}{\lambda_\rho}+\left(3+\frac{\bar{n}_{\rho-}}
{\bar{n}_\rho}\right)\frac{\dot{T}}{T}
 +\frac{1}{\tau}\notag\\
&=2\times\frac{1}{2}\langle\sigma_{\pi\pi\rightarrow\rho\rho}v_{\pi\pi}\rangle
  u^0\frac{\lambda^2_\pi \bar{n}^2_\pi}{\lambda_\rho \bar{n}_\rho}
 -2\times\frac{1}{2}\langle\sigma_{\rho\rho\rightarrow\pi\pi}v_{\rho\rho}
\rangle u^0\lambda_\rho\bar{n}_\rho \notag \\
&~~~+2\langle\sigma_{\pi K^{\ast}\rightarrow\rho K}v_{\pi K^\ast}\rangle 
  u^0\frac{\lambda_\pi \bar{n}_\pi\lambda_{K^\ast} \bar{n}_{K^\ast}}
{\lambda_\rho \bar{n}_\rho}
 -2\langle\sigma_{\rho K\rightarrow\pi K^{\ast}}v_{\rho K}\rangle 
u^0\lambda_K\bar{n}_K \notag \\
&~~~+2\langle\sigma_{\pi K\rightarrow\rho K^{\ast}}v_{\pi K}\rangle
  u^0\frac{\lambda_\pi \bar{n}_\pi\lambda_{K} \bar{n}_{K}}{\lambda_\rho 
\bar{n}_\rho}
 -2\langle\sigma_{\rho K^{\ast}\rightarrow\pi K}v_{\rho K^\ast}\rangle 
u^0\lambda_{K^\ast}\bar{n}_{K^\ast} \notag \\
&~~~+2\langle\sigma_{\pi K^{\ast}\rightarrow\rho K^{\ast}}v_{\pi K^{\ast}}
\rangle
  u^0\frac{\lambda_\pi \bar{n}_\pi\lambda_{K^\ast} \bar{n}_{K^\ast}}
{\lambda_\rho \bar{n}_\rho}
 -2\langle\sigma_{\rho K^{\ast}\rightarrow\pi K^{\ast}}v_{\rho K^{\ast}}
\rangle u^0\lambda_{K^\ast}\bar{n}_{K^\ast}\notag\\
&~~~+2\langle\sigma_{K\bar{K}^{\ast}\rightarrow\pi \rho}v_{K\bar{K}^{\ast}}
\rangle
  u^0\frac{\lambda_K \bar{n}_K\lambda_{K^\ast} \bar{n}_{K^\ast}}{\lambda_\rho 
\bar{n}_\rho}
 -2\langle\sigma_{\pi \rho\rightarrow K\bar{K}^{\ast}}v_{\pi \rho}\rangle 
u^0\lambda_\pi\bar{n}_\pi \notag \\
&~~~+2\langle\sigma_{K\bar{K}\rightarrow\rho\rho}v_{K\bar{K}}\rangle
  u^0\frac{\lambda^2_K \bar{n}^2_K}{\lambda_\rho \bar{n}_\rho}
 -2\times\frac{1}{2}\langle\sigma_{\rho\rho\rightarrow K\bar{K}}v_{\rho\rho}
\rangle u^0\lambda_\rho\bar{n}_\rho,
\end{align}
\begin{align}\label{eqk2}
&\frac{\dot{\lambda}_K}{\lambda_K}+\left(3+\frac{\bar{n}_{K-}}
{\bar{n}_K}\right)\frac{\dot{T}}{T}+\frac{1}{\tau}\notag\\
&=2\times\frac{1}{2}\langle\sigma_{K^{\ast} K^{\ast}\rightarrow KK}v_{K^{\ast} 
K^{\ast}}\rangle
  u^0\frac{\lambda^2_{K^\ast} \bar{n}^2_{K^\ast}}{\lambda_K \bar{n}_K}
 -2\times\frac{1}{2}\langle\sigma_{KK \rightarrow K^{\ast} K^{\ast}}
v_{KK}\rangle u^0\lambda_K\bar{n}_K \notag \\
&~~~+\frac{1}{2}\langle\sigma_{K^{\ast} K^{\ast}\rightarrow KK^{\ast}}
v_{K^{\ast} K^{\ast}}\rangle
  u^0\frac{\lambda^2_{K^\ast} \bar{n}^2_{K^\ast}}{\lambda_K \bar{n}_K}
 -\langle\sigma_{KK^{\ast} \rightarrow K^{\ast} K^{\ast}}v_{KK^{\ast}}\rangle 
u^0\lambda_{K^\ast}\bar{n}_{K^\ast} \notag \\
&~~~+\langle\sigma_{K^{\ast}\bar{K}^{\ast}\rightarrow K\bar{K}}v_{K^{\ast} 
\bar{K}^{\ast}}\rangle
  u^0\frac{\lambda^2_{K^\ast} \bar{n}^2_{K^\ast}}{\lambda_K \bar{n}_K}
 -\langle\sigma_{K\bar{K} \rightarrow K^{\ast}\bar{K}^{\ast}}v_{K\bar{K}}
\rangle u^0\lambda_K\bar{n}_K \notag \\
&~~~+\langle\sigma_{K^{\ast}\bar{K}^{\ast}\rightarrow K\bar{K}^{\ast}}
v_{K^{\ast}\bar{K}^{\ast}}\rangle
  u^0\frac{\lambda^2_{K^\ast} \bar{n}^2_{K^\ast}}{\lambda_K \bar{n}_K}
 -\langle\sigma_{K\bar{K}^{\ast}\rightarrow K^{\ast}\bar{K}^{\ast}}
v_{K\bar{K}^{\ast}}\rangle
  u^0\lambda_{K^\ast}\bar{n}_{K^\ast} \notag \\
&~~~+\langle\sigma_{\pi K^{\ast}\rightarrow\rho K}v_{\pi K^{\ast}}\rangle
  u^0\frac{\lambda_\pi\bar{n}_\pi\lambda_{K^\ast}\bar{n}_{K^\ast}}
{\lambda_K\bar{n}_K}
 -\langle\sigma_{\rho K\rightarrow\pi K^{\ast}}v_{\rho K}\rangle 
u^0\lambda_\rho\bar{n}_\rho \notag \\
&~~~+\langle\sigma_{\rho K^{\ast}\rightarrow\pi K}v_{\rho K^\ast}\rangle
  u^0\frac{\lambda_\rho\bar{n}_\rho\lambda_{K^\ast}\bar{n}_{K^\ast}}
{\lambda_K\bar{n}_K}
 -\langle\sigma_{\pi K\rightarrow\rho K^{\ast}}v_{\pi K}\rangle 
u^0\lambda_\pi\bar{n}_\pi \notag \\
&~~~+\langle\sigma_{\rho K^{\ast}\rightarrow\rho K}v_{\rho K^\ast}\rangle
  u^0\frac{\lambda_\rho\bar{n}_\rho\lambda_{K^\ast}\bar{n}_{K^\ast}}
{\lambda_K\bar{n}_K}
 -\langle\sigma_{\rho K\rightarrow\rho K^{\ast}}v_{\rho K}\rangle 
u^0\lambda_\rho\bar{n}_\rho \notag \\
&~~~+\frac{1}{2}\langle\sigma_{\pi\pi\rightarrow K\bar{K}}v_{\pi\pi}\rangle
  u^0\frac{\lambda^2_\pi \bar{n}^2_\pi}{\lambda_K \bar{n}_K}
 -\langle\sigma_{K\bar{K}\rightarrow\pi\pi}v_{K\bar{K}}\rangle 
u^0\lambda_K\bar{n}_K \notag \\
&~~~+\langle\sigma_{\pi \rho\rightarrow K\bar{K}^{\ast}}v_{\pi \rho}\rangle
  u^0\frac{\lambda_\pi\bar{n}_\pi\lambda_\rho\bar{n}_\rho}{\lambda_K\bar{n}_K}
 -\langle\sigma_{K\bar{K}^{\ast}\rightarrow\pi \rho}v_{K\bar{K}^{\ast}}\rangle
  u^0\lambda_{K^\ast}\bar{n}_{K^\ast} \notag \\
&~~~+\frac{1}{2}\langle\sigma_{\rho\rho\rightarrow K\bar{K}}v_{\rho\rho}\rangle
  u^0\frac{\lambda^2_\rho\bar{n}^2_\rho}{\lambda_K \bar{n}_K}
 -\langle\sigma_{K\bar{K}\rightarrow\rho\rho}v_{K\bar{K}}\rangle 
u^0\lambda_K\bar{n}_K,
\end{align}
\begin{align}\label{eqkstar2}
&\frac{\dot{\lambda}_{K^\ast}}{\lambda_{K^\ast}}+\left(3
+\frac{\bar{n}_{{K^\ast}-}}{\bar{n}_{K^\ast}}\right)
 \frac{\dot{T}}{T}+\frac{1}{\tau}\notag\\
&=2\times\frac{1}{2}\langle\sigma_{KK \rightarrow K^{\ast} K^{\ast}}
v_{KK}\rangle
  u^0\frac{\lambda^2_K \bar{n}^2_K}{\lambda_{K^\ast} \bar{n}_{K^\ast}}
 -2\times\frac{1}{2}\langle\sigma_{K^{\ast} K^{\ast}\rightarrow KK}v_{K^{\ast} 
K^{\ast}}\rangle
  u^0\lambda_{K^\ast}\bar{n}_{K^\ast} \notag \\
&~~~+\langle\sigma_{KK^{\ast} \rightarrow K^{\ast} K^{\ast}}
v_{KK^{\ast}}\rangle u^0\lambda_K\bar{n}_K
 -\frac{1}{2}\langle\sigma_{K^{\ast} K^{\ast}\rightarrow KK^{\ast}}v_{K^{\ast} 
K^{\ast}}\rangle
  u^0\lambda_{K^\ast}\bar{n}_{K^\ast} \notag \\
&~~~+\langle\sigma_{K\bar{K} \rightarrow K^{\ast}\bar{K}^{\ast}}v_{K\bar{K}}
\rangle
  u^0\frac{\lambda^2_K\bar{n}^2_K}{\lambda_{K^\ast}\bar{n}_{K^\ast}}
 -\langle\sigma_{K^{\ast}\bar{K}^{\ast}\rightarrow K\bar{K}}v_{K^{\ast} 
\bar{K}^{\ast}}\rangle
  u^0\lambda_{K^\ast}\bar{n}_{K^\ast} \notag \\
&~~~+\langle\sigma_{K\bar{K}^{\ast}\rightarrow K^{\ast}\bar{K}^{\ast}}
v_{K\bar{K}^{\ast}}\rangle
  u^0\lambda_K\bar{n}_K
 -\langle\sigma_{K^{\ast}\bar{K}^{\ast}\rightarrow K\bar{K}^{\ast}}
v_{K^{\ast}\bar{K}^{\ast}}\rangle
  u^0\lambda_{K^\ast}\bar{n}_{K^\ast} \notag \\
&~~~+\langle\sigma_{\rho K\rightarrow\pi K^{\ast}}v_{\rho K}\rangle
  u^0\frac{\lambda_\rho\bar{n}_\rho\lambda_K\bar{n}_K}{\lambda_{K^\ast}
\bar{n}_{K^\ast}}
 -\langle\sigma_{\pi K^{\ast}\rightarrow\rho K}v_{\pi K^{\ast}}\rangle
  u^0\lambda_\pi\bar{n}_\pi \notag \\
&~~~+\langle\sigma_{\pi K\rightarrow\rho K^{\ast}}v_{\pi K}\rangle
  u^0\frac{\lambda_\pi\bar{n}_\pi\lambda_K\bar{n}_K}{\lambda_{K^\ast}
\bar{n}_{K^\ast}}
 -\langle\sigma_{\rho K^{\ast}\rightarrow\pi K}v_{\rho K^\ast}\rangle
  u^0\lambda_\rho\bar{n}_\rho \notag \\
&~~~+\langle\sigma_{\rho K\rightarrow\rho K^{\ast}}v_{\rho K}\rangle
  u^0\frac{\lambda_\rho\bar{n}_\rho\lambda_K\bar{n}_K}{\lambda_{K^\ast}
\bar{n}_{K^\ast}}
-\langle\sigma_{\rho K^{\ast}\rightarrow\rho K}v_{\rho K^\ast}\rangle
  u^0\lambda_\rho\bar{n}_\rho \notag \\
&~~~+\langle\sigma_{\pi \rho\rightarrow K^{\ast}\bar{K}}v_{\pi \rho}\rangle
  u^0\frac{\lambda_\pi\bar{n}_\pi\lambda_\rho\bar{n}_\rho}{\lambda_{K^\ast}
\bar{n}_{K^\ast}}
 -\langle\sigma_{K^{\ast}\bar{K}\rightarrow\pi \rho}v_{K^{\ast}\bar{K}}\rangle 
u^0\lambda_K\bar{n}_K,
\end{align}
where the overdots denote the derivative with respect to the proper 
time $\tau$.

For the purpose of studying the role of quark-interchange processes,
it is enough to only consider the longitudinal expansion of hadronic matter.
The relativistic hydrodynamic equation is
\begin{equation}\label{1}
\partial_{\mu}T^{\mu\nu}=0,
\end{equation}
where $T^{\mu\nu}$ is the energy-momentum tensor given by
\begin{equation}\label{2}
T^{\mu\nu}=(\epsilon+P)u^{\mu}u^{\nu}-Pg^{\mu\nu},
\end{equation}
where $\epsilon$ is energy density and $P$ is pressure. The
simple form of $T^{\mu\nu}$ above holds for an ideal fluid where
viscosity effects have been neglected. The Bjorken's scaling
solution of the hydrodynamic equation is \cite{Bjo83}
\begin{equation}\label{Bjorken}
\frac{d \epsilon}{d \tau}+\frac{\epsilon+P}{\tau}=0.
\end{equation}
The energy density is 
\begin{equation}
\epsilon = \epsilon_\pi + \epsilon_\rho +\epsilon_K +\epsilon_{\bar K} 
+\epsilon_{K^\ast} +\epsilon_{\bar {K}^\ast}
\end{equation}
where $\epsilon_\pi$, $\epsilon_\rho$, $\epsilon_K$, $\epsilon_{\bar K}$, 
$\epsilon_{K^\ast}$ and $\epsilon_{\bar {K}^\ast}$ are the energy densities of 
$\pi$, $\rho$, $K$, $\bar K$, $K^\ast$ and $\bar {K}^\ast$, respectively.
In order to solve Eq. (\ref{Bjorken}), 
relation of pressure and energy density is needed. 
Detailed studies \cite{Sor95,Bra98,Kolb1,Kolb2} have
shown that the relation can be $P=0.15\epsilon$.
Once cross sections for the reactions concerned are known, Eqs.
(\ref{eqpi2})--(\ref{eqkstar2}) and (\ref{Bjorken}) determine time dependence
of $T(\tau)$, $\lambda_\pi(\tau)$, $\lambda_\rho(\tau)$,
$\lambda_K(\tau)$ and $\lambda_{K^\ast}(\tau)$.

\section{Cross sections for meson-meson reactions}
The meson-meson cross sections entailed 
in the master rate equations in Section 2 
are the isospin-averaged cross sections that are obtained by taking
the average over the isospin states of the two initial mesons and the sum over
the isospin states of the two final mesons
\begin{equation}
\sigma_{ij\to i^\prime j^\prime}(\sqrt{s})
=\frac{1}{(2I_1+1)(2I_2+1)}\sum_{I}(2I+1)\sigma (I,\sqrt{s}),
\end{equation}
where $I_1$ and $I_2$ are the isospins of the two initial mesons, respectively,
and $\sigma (I,\sqrt{s})$ is the spin-averaged cross section for the reaction
with the total isospin $I$. 

We explicitly decompose the cross section $\sigma (I,\sqrt {s})$ into three
parts: the first is $\sigma_{\rm qi}$
from the quark-interchange process, the second $\sigma_{\rm anni}$
from the annihilation processes and the third $\sigma_{\rm res}$
from the resonant processes. Since the momenta and the coordinates of the quark
and antiquark constituents of the
final mesons in the quark-interchange process $A(q_1\bar {q}_1) +
B(q_2\bar {q}_2) \to C(q_1\bar {q}_2) + D(q_2\bar {q}_1)$ are different from
those in the annihilation processes or the resonant processes 
$A(q_1\bar {q}_1) + B(q_2\bar {q}_2) \to 
C(q_1\bar {q}_1) + D(q_2\bar {q}_2)$, there is no interference between the
quark-interchange process and the annihilation processes and between the
quark-interchange process and the resonant processes. No interference of the 
annihilation processes and the resonant processes is usually assumed. 
Then the cross section for a reaction is written as
\begin{equation}\label{deccs}
\sigma (I,\sqrt{s}) =c_{\rm qi} \sigma_{\rm qi}(I,\sqrt{s}) 
+ c_{\rm anni} \sigma_{\rm anni} (I,\sqrt{s}) 
+ c_{\rm res} \sigma_{\rm res}(I,\sqrt{s}).
\end{equation}
The coefficients $c_{\rm qi}$, $c_{\rm anni}$ and $c_{\rm res}$ that take 
values of 0 or 1 are listed in 
Table~\ref{vabc}. In the following three subsections $\sigma_{\rm qi}$, 
$\sigma_{\rm anni}$ and $\sigma_{\rm res}$ are presented. For two isospin
channels all the three processes contribute. For most isospin channels only
the quark-interchange processes or the annihilation processes contribute.
Quark-interchange-induced reactions refer to the channels where only the
quark interchange works. The quark-interchange  processes include the 
quark-interchange-induced reactions.

\subsection{Cross sections for quark-interchange processes}
In Ref. \cite{Li07} we have obtained unpolarized cross sections for some
meson-meson nonresonant reactions governed only by the quark-interchange
mechanism. The cross sections rely on mesonic quark-antiquark wave functions 
and constituent-constituent interaction.
The quark-antiquark relative-motion wave functions are determined 
by the Buchm\"{u}ller-Tye potential \cite{Buc81} that arises from 
color confinement and one-gluon exchange plus one- and two-loop corrections.
The constituent-constituent interaction includes the Buchm\"{u}ller-Tye 
potential what is nonrelativistic, central and spin-independent, and spin-spin
terms that are obtained by performing Foldy-Wouthuysen canonical 
transformations to a relativistic two-constituent Hamiltonian that includes 
the linear confinement and a relativistic one-gluon-exchange potential plus 
perturbative one- and two-loop corrections \cite{Xu02}. The wave functions and
the interaction can reproduce the experimental mass splittings between the 
ground-state pseudoscalar octet mesons and the ground-state vector nonet mesons
\cite{Xu02}. The wave functions and the interaction were employed \cite{Li07}
to calculate cross sections for nonresonant reactions
$A(q_1\bar {q}_1)+B(q_2\bar {q}_2) \to C(q_1\bar {q}_2)+D(q_2\bar {q}_1)$ which
are endothermic or exothermic. The $\sqrt s$-dependence of numerical cross 
sections exhibited in Ref. \cite{Li07}
show a peak for each of endothermic reactions and large magnitudes very near
the threshold energies of exothermic reactions. 
For convenient use of the numerical cross sections in solving the master rate 
equations, parametrizations similar to Ref. \cite{Bar03} read
\begin{equation}\label{finterch}
\sigma_{\rm qi}(I,\sqrt{s})=\sigma_{\rm
max}\left(\frac{\epsilon}{\epsilon_{\rm endo}}\right)^a
\exp\left[a\left(1-\frac{\epsilon}{\epsilon_{\rm
endo}}\right)\right],
\end{equation}
where $\epsilon=\sqrt{s}-\sqrt{s_0}$ shows difference from the threshold energy
$\sqrt{s_0}$.
Values for the parameters $\sigma_{\rm max}$, $\epsilon_{\rm endo}$ and
$a$ for the quark-interchange-induced endothermic
reactions given in the introduction are shown in Table~\ref{t_interch}.

The right-hand side of Eq. (22) indicates the zero value of cross section at
the threshold energy and a fall of cross section at $\sqrt {s} \to \infty$.
$\epsilon_{\rm endo}$ is close to the center-of-mass energy at which an
endothermic reaction reaches maximum cross section $\sigma_{\rm max}$. The 
power function and the exponential function do leave a curve that is not
symmetric with respect to the peak. The peak is in the energy region that
is accessible to meson-meson reactions in mesonic matter.

Table 2 shows the channels with the highest isospins. Since quark-interchange
processes can also take place in low isospin channels as seen in Table 1,
flavor matrix elements $f_{\rm flavor}(I)$ of the quark-interchange processes 
are listed in Table \ref{f_matrix} for different isospins.
The entry $f_{\rm flavor}(1)=0$ for $\pi \pi \rightarrow \rho \rho$ for $I=1$ 
means that the reaction in $I=1$ is
forbidden. This is consistent with the fact that the antisymmetric state of
$\pi \pi$ is not allowed.
The discrepancy of the cross sections for different
isospin channels of a reaction results from the
different flavor matrix elements while matrix elements involving spin
and spatial wave functions are equal. Then, for instance, we have
$\sigma_{\rm qi}(I=0, \sqrt {s})=\frac {1}{4}\sigma_{\rm qi}(I=2, \sqrt {s})$
and $\sigma_{\rm qi}(I=1, \sqrt {s})=0$ for $\pi \pi \rightarrow \rho \rho$. 
Let $I_{\rm max}$ denote the highest isospin of a reaction. 
The cross section for a channel listed in Table 3 is 
\begin{equation}
\sigma_{\rm qi}(I, \sqrt {s})=\frac {f^2_{\rm flavor}(I)}
{f^2_{\rm flavor}(I_{\rm max})}\sigma_{\rm qi}(I_{\rm max}, \sqrt {s}).
\end{equation}

The cross section for an exothermic reaction $i^\prime (S_3 I_3) + j^\prime 
(S_4 I_4) \to i (S_1 I_1) + j (S_2 I_2)$ is 
obtained from the endothermic reaction $i j \to i^\prime j^\prime$
by the detailed balance
\begin{equation}\label{balance}
\sigma_{i^\prime j^\prime \rightarrow ij}=\frac{\vec{P}^2}
{\vec{P}^{\prime^2}}
\frac{g_{\rm i}}{g_{\rm f}} \sigma_{ij \rightarrow i^\prime j^\prime},
\end{equation}
where $g_{\rm i}=(2S_1+1)(2I_1+1)(2S_2+1)(2I_2+1)$ and $g_{\rm
f}=(2S_3+1)(2I_3+1)(2S_4+1)(2I_4+1)$ denote the spin-isospin
degeneracy factors of initial particles with the spins $S_1$ and
$S_2$ as well as the isospins $I_1$ and $I_2$, and of final
particles with the spins $S_3$ and $S_4$ as well as the isospins
$I_3$ and $I_4$, respectively; $\vec{P}$ and $\vec{P}^\prime$ denote momenta of
an initial meson and a final meson in the center-of-momentum frame of the
reaction $ij \to i^\prime j^\prime$, respectively. 

\subsection{Cross sections for annihilation processes}
The isospin-averaged cross section for an annihilation reaction
can be parametrized by \cite{Bro91,Cas97}
\begin{equation}\label{fanni}
\sigma_{\rm anni}(\sqrt{s})=b\left(1-\frac{s_0}{s}\right)^c.
\end{equation}
Values of the parameter $b$ and the dimensionless parameter $c$ for the
reactions $\pi \pi \rightarrow K \bar{K} $, $\pi \rho \rightarrow K
\bar{K}^\ast(K^\ast \bar{K})$ and $K \bar{K} \rightarrow \rho \rho$
are given in Table~\ref{table_anni}. Since no experimental data are
available, we assume that cross sections for other annihilation processes 
listed in Table 1 possess Eq. (25) with the same parameters $b$ and $c$
as the reaction $\pi \pi \rightarrow K \bar{K}$. The treatment of the
annihilation processes is simple.

\subsection{Cross sections for resonant processes}
Cross sections for resonant processes are generally described by the 
Breit-Wigner formula \cite{RQMD1,RQMD2,HSD,UrQMD1,UrQMD2}
\begin{equation}\label{fres}
\sigma_{\rm res}(\sqrt{s})=\frac{2J+1}{(2S_1+1)(2S_2+1)}\frac{\pi}{\vec{P}^2}
\frac{\Gamma^2 B_{\rm{in}}B_{\rm{out}}}{(\sqrt{s}-m_{\rm
R})^2+\Gamma^2/4},
\end{equation}
where a resonance has its spin $J$, its energy $m_{\rm R}$ and its full width
$\Gamma$, and $B_{\rm{in}}$ and
$B_{\rm{out}}$ are the branching fractions of the resonance decays into the
initial state and the final state, respectively. We take account of the 
resonances $f_0(1370)$, $\rho (1450)$, $f_0(1500)$ and $\rho_3(1690)$ for
$\pi \pi \leftrightarrow \rho \rho$, $K_1(1270)$, $K_1(1400)$, $K^\ast(1410)$, 
$K_2^\ast(1430)$, $K^*(1680)$ and $K_3^*(1780)$ for $\pi K^\ast 
\leftrightarrow \rho K$, $f_0(980)$, $\phi(1020)$, $f_2(1270)$, $f_0(1370)$,
$\rho(1450)$, $f_0(1500)$, $f_2^\prime(1525)$, $\rho_3(1690)$, $\rho(1700)$,
$f_0(1710)$, $f_2(1810)$ and $f_4(2050)$ for $\pi \pi \leftrightarrow K\bar K$.
Since the full widths and 
the branching fractions for some resonances have not been fixed by measurements
\cite{PDG}, their values we select for the reactions 
$\pi\pi \leftrightarrow \rho\rho$, $\pi K^\ast \leftrightarrow \rho K$ 
and $\pi\pi \leftrightarrow K \bar{K}$ are listed in Table \ref{rhorho}.

\section{Results and discussions}
In this section we represent and discuss results that are from solving the
master rate equations in combination with the hydrodynamic
equation (18) simultaneously by numerical integration using a fourth
order Runge-Kutta method. The inelastic 2-to-2 scattering in the master rate 
equations includes the three types of processes:
the quark-interchange processes, the annihilation processes
and the resonant processes. As a good approximation,
we assume that hadronization of quark-gluon plasma at the critical temperature 
$T_{\rm c}=175$ MeV \cite{Kar01} only produces $\pi$, $\rho$, $K$, $\bar K$,
$K^\ast$ and $\bar {K}^\ast$.
We assume that the hadronization is finished at 
$\tau_{\rm h}=5.6~{\rm fm}/c$ and at the moment mesonic matter has 
$\lambda_{\pi}=0.7$, 
$\lambda_{\rho}=0.7$, $\lambda_{K}=0.5$ and $\lambda_{K^{\ast}}=0.2$ for the
fugacities of $\pi$, $\rho$, $K$ and $K^\ast$, respectively. 
We consider mesonic matter at or near mid-rapidity, i.e., $u^\mu \approx 
(1,0,0,0)$, solve Eqs. (12)-(15) and (18), and terminate numerical 
calculations when mesonic matter reaches the kinetic freeze-out
temperature $T_{\rm fz}=105~\rm{MeV}$, which corresponds to a freeze-out time
of the order of 30 fm/$c$.

If the inelastic 2-to-2 scattering is switched off, ie., the source 
terms are zero, the master rate equations become
\begin{equation}
 \partial_{\mu}(n_i u^{\mu})=0,
\end{equation}
which have the solutions
\begin{equation}
n_i \sim \frac {1}{\tau}.
\end{equation}
Together with Eq. (8) we have
\begin{equation}
\lambda_i \sim \frac {1}{u^0 \bar {n}_i \tau}.
\end{equation}
For massive bosons, $\bar {n}_i$ given by Eq. (9) and energy densities 
do not simply rely on a power of
$T$, and the hydrodynamic equation cannot guarantee $\bar {n}_i \tau$ as 
constants. Therefore, $\lambda_i$ depend on the proper time $\tau$ unlike the
case of massless bosons which fugacities are constants. Fugacities for $\pi$,
$\rho$, $K$ and $K^\ast$ are denoted by $\lambda_{\pi \rm no}$, 
$\lambda_{\rho \rm no}$, $\lambda_{K \rm no}$ and $\lambda_{K^* \rm no}$,
respectively, and are plotted as solid curves in Fig. 1. The solid curves
for $\rho$, $K$ and $K^\ast$ rise  with the increase of time. The fugacity of 
the lightest meson first decreases slightly and then increases. Compared to 
the results in the absence of the source terms,
we show meson fugacities by the dashed curves and indicate the meson 
fugacities by $\lambda_{i\rm qar}$ ($i=\pi, \rho, K, K^\ast$)
while the quark-interchange processes, 
the annihilation processes and the resonant processes are all included. 
The differences between $\lambda_{i\rm qar}$ and $\lambda_{i\rm no}$
due to the inelastic 2-to-2 scattering are obvious. To
show how the quark-interchange processes modify fugacities, we show the 
fugacities by dotted curves and denote the fugacities by $\lambda_{i\rm qi}$
($i=\pi, \rho, K, K^\ast$) while only the quark-interchange 
processes govern time dependence of the fugacities. In most of the range
$5.6~{\rm fm}/c < \tau < 30~{\rm fm}/c$ the absolute values of
the fugacity differences, $\mid 
\lambda_{i\rm qi} - \lambda_{i\rm no} \mid$, caused by the 
quark-interchange processes are smaller than $\mid \lambda_{i\rm qar} - 
\lambda_{i\rm no} \mid$ caused by the three types of processes.

To show a role of the quark-interchange processes, we define
\begin{equation}
R_i = \frac {\lambda_{i\rm qi}- \lambda_{i\rm no} }
{ \lambda_{i\rm qar}- \lambda_{i\rm no} }.
\end{equation}
The larger the absolute values of $R_i$,
the more important the quark-interchange processes.  If $R_i>0$, either of 
the quark-interchange processes and the combination of the three 
types of processes increases (reduces) fugacities relative to 
$\lambda_{i\rm no}$. If $R_i<0$, the quark-interchange processes
increase (reduce) fugacities relative to $\lambda_{i\rm no}$
while the annihilation and resonant processes reduce (increase) fugacities.
Values of $R_{i}$ change with the increase of time. At 
$\tau =20~{\rm fm}/c$, $R_{\pi}= 0.37$, 
$R_{\rho}= 0.30$, $R_K= 0.21$ and $R_{K^*}= 0.27$.

To quantitatively determine the importance of the quark-interchange processes,
we define the average of the absolute value of $R_i$ by
\begin{equation}
\bar {R}_i = \frac {\int_{\tau_{\rm h}}^{\tau_{\rm fz}} d\tau \mid R_i \mid}
{\tau_{\rm fz} - \tau_{\rm h} }.
\end{equation}
where $\tau_{\rm fz}$ is the freeze-out time of mesonic matter.
Then, $\bar {R}_{\pi} =0.53$, $\bar {R}_{\rho} =0.30$, $\bar {R}_{K} =0.21$ 
and $\bar {R}_{K^*} =0.27$. Hence, with the set of initial fugacities,
$\lambda_\pi=0.7$, $\lambda_\rho=0.7$, $\lambda_K=0.5$ and $\lambda_{K^*}=0.2$,
for the master rate equations and the hydrodynamic equation,
the quark-interchange processes are important in the contribution of
the inelastic 2-to-2 scattering to the evolution of mesonic matter.
But the conclusion is only limited to this set. We need to examine
${\bar R}_i$ versus other initial fugacities. 
Initial fugacities of mesonic matter depend on incident energies in 
nucleus-nucleus collisions. The Au-Au 
collisions have been carried out at various energies of per pair of colliding 
nucleons allowed by the Relativistic Heavy Ion Collider (RHIC)
and Pb-Pb collisions at the Large Hadron Collider (LHC) have been
performed at higher energies. Different nucleus-nucleus collisions at
different energies produce mesonic matter with different magnitudes of initial
fugacities. Therefore, we use a wide range of initial fugacities to check the 
importance of the quark-interchange processes. It is impossible to plot graphs
for $\bar {R}_i$ ($i=\pi$, $\rho$, $K$, $K^\ast$) versus the four variables
$\lambda_\pi$, $\lambda_\rho$, $\lambda_K$ and $\lambda_{K^\ast}$, but we can 
tabulate $\bar {R}_i$ ($i=\pi$, $\rho$, $K$, $K^\ast$) versus a finite sets of
initial fugacities in the range between 0 and 1. We take 81 sets in which 
$\lambda_\pi=0.35, 0.65, 0.95$, $\lambda_\rho =0.15, 0.45, 0.75$, $\lambda_K=
0.25, 0.55, 0.85$ and $\lambda_{K^\ast}=0.15, 0.55, 0.95$. Statistically, the 
81 sets can tell us how important the quark-interchange processes are. The
freeze-out time $\tau_{\rm fz}$ depends on initial fugacities. Averages
$\bar {R}_i$ ($i=\pi$, $\rho$, $K$, $K^\ast$) for the 81 sets of initial
fugacities are listed in the middle four columns in Tables 6-8.

The average may be as large as 33.07 and this occurs when the contributions 
of the quark-interchange, annihilation and resonant processes can cancel each 
other. The average may be as small as 0.01. Most of the entries in the middle
four columns are above 0.2. Therefore, the quark-interchange processes are 
important in the contribution of the inelastic 2-to-2 scattering to the
evolution of mesonic matter.

Via HIJING Monte Carlo simulations \cite{HIJING1,HIJING2,HIJING3}, initial 
fugacities of
deconfined gluons and quarks produced in central Au-Au collisions at
$\sqrt {s_{NN}}=200$ GeV are about 0.2 and 0.032, respectively, as seen in
the second set of initial conditions of quark-gluon plasma
in Table I of Ref. \cite{Xu96}. Using the 
ratio 0.2 of strange-quark number to up-quark number \cite{Biro99}, 
that may reproduce the measured ratios $K^-/\pi^-=0.15 \pm 0.02$ and
$K^{*0}/K^-=0.205 \pm 0.033$ at midrapidity \cite{Baran}, we solve master rate 
equations of quark-gluon plasma given in Ref. \cite{Lev95} to obtain time
dependence of fugacities of gluons and quarks. We obtain the time 
$\tau_{\rm h} \approx 5.6$ fm/$c$, the fugacities $\lambda_g \approx 0.75$ and 
$\lambda_q \approx 0.52$ at $T_{\rm c}$. Coalescence of quarks 
and antiquarks forms mesons at $T_{\rm c}$. Assume that the formed mesons are
only $\pi$, $\rho$, $K$, $\bar K$, $K^*$ and $\bar {K}^*$ and that 
$\lambda_\pi = \lambda_\rho$ and $\lambda_K = \lambda_{K^*}$ at $T_{\rm c}$. 
Then we obtain  $\lambda_\pi = \lambda_\rho \approx 1.31$ and 
$\lambda_K = \lambda_{K^*} \approx 0.56$ that are the initial fugacities 
of mesonic matter produced in central Au-Au collisions at 
$\sqrt {s_{NN}}=200$ GeV. For other nucleus-nucleus collisions 
meson fugacities are different from these values. The pion fugacity may 
not equal the rho fugacity at $T_{\rm c}$ and the kaon fugacity may not equal 
the vector kaon fugacity. If $\sqrt {s_{NN}}$ decreases continuously, meson 
fugacities decrease continuously.
Therefore, $\lambda_\pi = \lambda_\rho = 0.7$, $\lambda_K = 0.5$ and 
$\lambda_{K^*} = 0.2$ used in this section and
some sets of initial fugacities listed in Tables 6-8 can be covered 
in some nucleus-nucleus collisions at some values of $\sqrt {s_{NN}}$. 
Since the strange-quark number is less than half the up-quark number, 
those sets of initial
fugacities in which both $\lambda_K$ and $\lambda_{K^*}$ are larger than 
$\lambda_\pi$ and $\lambda_\rho$ are not possible. But such sets of 
initial fugacities yield $\bar {R}_i>0.2$ ($i=\pi, \rho, K, K^*$) in more than
half the entries of $\bar {R}_i$. Therefore, 
the impossible sets of initial fugacities help us more firmly establish that 
the quark-interchange processes are important in the contribution of the 
inelastic 2-to-2 scattering to the evolution of mesonic matter.
  
It is shown in Table 1 that fourteen reaction channels involve the 
quark-interchange processes, sixteen channels the annihilation 
processes and three channels the resonant processes.
In a reaction where a quark-interchange process
occurs, the channel with the highest isospin is only induced by the
quark-interchange process and the number $2I+1$ of isospin component in Eq.
(20) can enhance the capability of the process in influencing the evolution of 
mesonic matter.

\section{Results pertinent to $2 \leftrightarrow 1$ processes}
Since we only consider $\pi$, $\rho$, $K$, $\bar K$, $K^*$ and 
$\bar {K}^*$, resonances involved are $\rho$, $K^*$ and $\bar {K}^*$. Then we 
include $\pi \pi \leftrightarrow \rho$, $\pi K \leftrightarrow K^*$
and $\pi \bar {K} \leftrightarrow \bar {K}^*$. Let $\Gamma_{\rho \to \pi \pi}$,
$\Gamma_{K^* \to \pi K}$ and $\Gamma_{\bar {K}^* \to \pi \bar K}$ 
be the decay widths of $\rho \to \pi \pi$, $K^* \to \pi K$ and $\bar {K}^* \to
\pi \bar K$, respectively. We add the following four expressions
\begin{displaymath}
2\Gamma_{\rho \to \pi \pi}n_\rho - 2 \times \frac {1}{2} \langle 
\sigma_{\pi \pi \to \rho} v_{\pi \pi} \rangle n_\pi n_\pi
+\Gamma_{K^\ast \to \pi K} n_{K^\ast} 
+\Gamma_{\bar {K}^\ast \to \pi \bar K} n_{\bar {K}^\ast} 
\end{displaymath}
\begin{displaymath}
- \langle \sigma_{\pi K \to K^\ast} v_{\pi K} \rangle n_\pi n_K
- \langle \sigma_{\pi \bar {K} \to \bar {K}^\ast} v_{\pi \bar K} \rangle 
n_\pi n_{\bar K},
\end{displaymath}
\begin{displaymath}
-\Gamma_{\rho \to \pi \pi}n_\rho + \frac {1}{2} \langle 
\sigma_{\pi \pi \to \rho} v_{\pi \pi} \rangle n_\pi n_\pi,
\end{displaymath}
\begin{displaymath}
\Gamma_{K^\ast \to \pi K}n_{K^\ast} - \langle 
\sigma_{\pi K \to K^\ast} v_{\pi K} \rangle n_\pi n_K,
\end{displaymath}
and
\begin{displaymath}
-\Gamma_{K^\ast \to \pi K}n_{K^\ast} + \langle 
\sigma_{\pi K \to K^\ast} v_{\pi K} \rangle n_\pi n_K,
\end{displaymath}
to the source terms $\Psi_\pi$, $\Psi_\rho$, $\Psi_K$ and $\Psi_{K^\ast}$ in
Eqs. (2)-(5), respectively, to establish master rate equations with the
$2 \leftrightarrow 1$ mesonic processes. 
Obtained from the experimental data \cite{Fla84},
the cross section for $\pi \pi \to \rho$ is
\begin{equation}
\sigma_{\pi \pi \to \rho} = \frac {80~{\rm mb}}
{1+4(\sqrt {s}-m_\rho)^2/\Gamma^2_{\rho \to \pi \pi}}.
\end{equation}
The cross section for $\pi K \to K^\ast$ or $\pi \bar {K} \to \bar {K}^\ast$
can be found in Refs. \cite{ART1,ART2}. From solutions of the master rate 
equations with the $2 \leftrightarrow 1$ processes
and the hydrodynamic equation (18), we get the average values $\bar {R}_i$ at
various initial fugacities and list them in the right four columns in Tables 
6-8.

In the tables about 96\% of the entries in the right four columns have values 
larger than 0.5 and 53\% larger than 1. Moreover, most of $\bar {R}_i$ ($i=\pi,
\rho, K, K^*$) are larger than the corresponding ones derived from the master
rate equations without the $2 \leftrightarrow 1$ processes.
Therefore, while the quark-antiquark annihilation processes and the resonant
processes are taken into account, the quark-interchange processes must be 
included on an equal footing.

\section{Results pertinent to transverse expansion}
In the preceding sections we have considered only the longitudinal 
expansion for mesonic matter. In this section we rely on
both longitudinal and transverse expansion to deal with 
mesonic matter produced in central collisions. The four-velocity of the
local reference frame is $u^\mu=\gamma (\frac {t}{\tau},v_r,0,\frac {z}{\tau})$
with $\gamma =1/\sqrt {1-v^2_r}$, where $v_r$ is the transverse velocity. The
left-hand side in Eq. (1) becomes
\begin{equation}
\partial_\mu(n_i u^\mu)=\gamma \frac {\partial n_i}{\partial \tau}
+n_i (\frac {\partial \gamma}{\partial \tau} + \frac {\gamma}{\tau})
+\frac {1}{r} \frac {\partial}{\partial r}(rn_i\gamma v_r).
\end{equation}
For matter uniformly distributed,
\begin{equation}
\partial_\mu(n_i u^\mu)=\gamma \frac {\partial n_i}{\partial \tau}
+n_i \gamma^3 v_r \frac {\partial v_r}{\partial \tau}
+n_i \gamma (2\gamma^2 v_r^2 +1) \frac {\partial v_r}{\partial r}
+\frac {n_i \gamma}{\tau} + \frac {n_i\gamma v_r}{r}.
\end{equation}
In terms of fugacities,
\begin{equation}
\partial_\mu(n_i u^\mu)=
u^0 \bar {n}_i\gamma \frac {\partial \lambda_i}{\partial \tau}
+u^0 \lambda_i \gamma \frac {\partial \bar {n}_i}{\partial T}
\frac {\partial T}{\partial \tau}
+2n_i \gamma^3 v_r \frac {\partial v_r}{\partial \tau}
+n_i \gamma (2\gamma^2 v_r^2 +1) \frac {\partial v_r}{\partial r}
+\frac {n_i \gamma}{\tau} + \frac {n_i\gamma v_r}{r}.
\end{equation}
Solving the master rate equations with Eq. (35) and
the $2 \leftrightarrow 1$ mesonic
processes and the hydrodynamic equations describing the longitudinal and 
transverse expansion in Ref. \cite{Ger86}, we obtain, for example, at 
$\lambda_\pi=0.7$, 
$\lambda_\rho=0.7$, $\lambda_K=0.5$ and $\lambda_{K^*}=0.2$, the average values
$\bar {R}_i$ ($i=\pi, \rho, K, K^*$) are 1.11, 0.78, 1.21 and 0.85, 
respectively. Corresponding to most sets of the initial fugacities 
listed in Tables 6-8, $\bar {R}_{\pi}$, $\bar {R}_{\rho}$, $\bar {R}_K$ 
and $\bar {R}_{K^*}$ are larger than 1. Therefore, we
must use the quark-interchange processes on an equal footing while the
annihilation processes and the resonant processes are considered.

\section{Summary}
We have established a set of master rate equations that describe time 
dependence of fugacities of pions, rhos, kaons and vector kaons in mesonic
matter. A meson-meson reaction is comprised of the quark-interchange process,
the annihilation processes and the resonant processes. The cross sections for 
the quark-interchange-induced reactions, that were obtained from the 
Buchm\"uller-Tye potential plus the spin-spin interaction, are parametrized 
for convenient use in studying the evolution of mesonic matter.

The variations of
fugacities of pions, rhos, kaons and vector kaons are governed by the
inelastic meson-meson scattering, the $2 \leftrightarrow 1$ mesonic processes
and the expansion of mesonic matter. In most reactions the
quark-interchange processes take place. If the number density of $\pi$ is
increased (reduced) by the quark-interchange processes, the number density of
$\rho$ is reduced (increased) in the same amount. This relation also holds true
for $K$ and $K^\ast$. Numerical results of the master rate equations show that 
the quark-interchange processes are important in the contribution of the 
inelastic 2-to-2 scattering to the evolution of mesonic matter.

\section*{Acknowledgments}
This work was supported by National Natural Science Foundation
of China under Grant No.~10675079.


\newpage
\begin{table}[!ht]
\tabcolsep 0pt 
\caption{Values of $c_{\rm qi}$, $c_{\rm anni}$ and $c_{\rm res}$.}
\label{vabc} 
\vspace*{-12pt}
\begin{center}
\def\temptablewidth{0.7\textwidth}
{\rule{\temptablewidth}{1pt}}
\begin{tabular*}{\temptablewidth}{@{\extracolsep{\fill}}llll}
Channel & $c_{\rm qi}$ & $c_{\rm anni}$ & $c_{\rm res}$  \\   
\hline
$I=2~\pi \pi \leftrightarrow \rho \rho $  &   1  & 0 & 0 \\
$I=1~\pi \pi \leftrightarrow \rho \rho $  &   0  & 1 & 0 \\
$I=0~\pi \pi \leftrightarrow \rho \rho $  &   1  & 1 & 1 \\
$I=1~K K \leftrightarrow K^\ast K^\ast $  &  1  & 0 & 0 \\
$I=0~K K \leftrightarrow K^\ast K^\ast $  &  1  & 0 & 0 \\
$I=1~K K^\ast \leftrightarrow K^\ast K^\ast $  &  1  & 0 & 0 \\
$I=0~K K^\ast \leftrightarrow K^\ast K^\ast $  &  1  & 0 & 0 \\
$I=1~K \bar {K} \leftrightarrow K^\ast \bar {K}^\ast $  &  0  & 1 & 0 \\
$I=0~K \bar {K} \leftrightarrow K^\ast \bar {K}^\ast $  &  0  & 1 & 0 \\
$I=1~K \bar {K}^\ast \leftrightarrow K^\ast \bar {K}^\ast $  &  0  & 1 & 0 \\
$I=0~K \bar {K}^\ast \leftrightarrow K^\ast \bar {K}^\ast $  &  0  & 1 & 0 \\
$I=3/2~\pi K^\ast \leftrightarrow \rho K$  &  1  & 0 & 0 \\
$I=1/2~\pi K^\ast \leftrightarrow \rho K$  &  1  & 1 & 1 \\
$I=3/2~\pi K \leftrightarrow \rho K^\ast$  &  1  & 0 & 0 \\
$I=1/2~\pi K \leftrightarrow \rho K^\ast$  &  1  & 1 & 0 \\
$I=3/2~\pi K^\ast \leftrightarrow \rho K^\ast$  &  1  & 0 & 0 \\
$I=1/2~\pi K^\ast \leftrightarrow \rho K^\ast$  &  1  & 1 & 0 \\
$I=3/2~\rho K \leftrightarrow \rho K^\ast$  &  1  & 0 & 0 \\
$I=1/2~\rho K \leftrightarrow \rho K^\ast$  &  1  & 1 & 0 \\
$I=1~\pi \pi \leftrightarrow K \bar K$  &  0  & 1 & 0 \\
$I=0~\pi \pi \leftrightarrow K \bar K$  &  0  & 1 & 1 \\
$I=1~\pi \rho \leftrightarrow K \bar {K}^\ast$  &  0  & 1 & 0 \\
$I=0~\pi \rho \leftrightarrow K \bar {K}^\ast$  &  0  & 1 & 0 \\
$I=1~K \bar K \leftrightarrow \rho \rho$  &  0  & 1 & 0 \\
$I=0~K \bar K \leftrightarrow \rho \rho$  &  0  & 1 & 0 \\
\end{tabular*}
{\rule{\temptablewidth}{1pt}}
\end{center}
\end{table}

\begin{table}[!ht]
\tabcolsep 0pt 
\caption{Parameters in Eq. (22).}
\label{t_interch} 
\vspace*{-12pt}
\begin{center}
\def\temptablewidth{0.7\textwidth}
{\rule{\temptablewidth}{1pt}}
\begin{tabular*}{\temptablewidth}{@{\extracolsep{\fill}}llll}
Channel & $\sigma_{\rm max}~(\rm{mb})$ & $\epsilon_{\rm endo}~(\rm{GeV})$ &
$a$  \\   \hline
$I=2~\pi \pi \to \rho \rho $  &   0.49991  & 0.20909 & 0.87446 \\
$I=1~K K \to K^\ast K^\ast $  &  0.61622  & 0.16539 & 0.4883\\
$I=1~K K^\ast \to K^\ast K^\ast $  &  0.85168  & 0.26399 & 1.05175\\
$I=3/2~\pi K^\ast \to \rho K$  &  1.40233  & 0.15023 & 1.07478\\
$I=3/2~\pi K \to \rho K^\ast$  &  0.49839  & 0.12056  & 0.40939\\
$I=3/2~\pi K^\ast \to \rho K^\ast$  &  0.49  & 0.21 & 0.88\\
$I=3/2~\rho K \to \rho K^\ast $  &  0.5081  & 0.3166 & 1.89693
\end{tabular*}
{\rule{\temptablewidth}{1pt}}
\end{center}
\end{table}

\begin{table}[!ht]
\tabcolsep 0pt \caption{Flavor matrix elements $f_{\rm flavor}(I)$.}
\label{f_matrix}
\vspace*{-12pt}
\begin{center}
\def\temptablewidth{0.7\textwidth}
{\rule{\temptablewidth}{1pt}}
\begin{tabular*}{\temptablewidth}{@{\extracolsep{\fill}}lccccc}
  & $I=0$ & $I=1$ & $I=2$ & $I=1/2$ & $I=3/2$ \\   \hline
$\pi \pi \to \rho \rho $ & $-1/2$ & $0$ & $1$ & &\\
$K K \to K^\ast K^\ast $ & $1$ & $1$ & &\\
$K K^\ast \to K^\ast K^\ast $ & $1$ & $1$ & &\\
$\pi K^\ast \to \rho K$  & & & & $-1/2$& $1$ \\
$\pi K \to \rho K^\ast$  & & & & $-1/2$& $1$ \\
$\pi K^\ast \to \rho K^\ast$  & & & & $-1/2$& $1$ \\
$\rho K \to \rho K^\ast $  & & & & $-1/2$& $1$
\end{tabular*}
{\rule{\temptablewidth}{1pt}}
\end{center}
\end{table}

\begin{table}[!ht]
\tabcolsep 0pt \caption{Values of $b$ and $c$.}
\label{table_anni} \vspace*{-12pt}
\begin{center}
\def\temptablewidth{0.5\textwidth}
{\rule{\temptablewidth}{1pt}}
\begin{tabular*}{\temptablewidth}{@{\extracolsep{\fill}}lll}
Reaction & $b~(\rm{mb})$ & $c$  \\
\hline
$\pi \pi \rightarrow K \bar{K} $  &  2.7  & 0.76  \\
$\pi \rho \rightarrow K \bar{K}^\ast(K^\ast \bar{K})$  &  0.4 & 0.5 \\
$K \bar{K} \rightarrow \rho \rho$  &  3.5  & 0.38
\end{tabular*}
{\rule{\temptablewidth}{1pt}}
\end{center}
\end{table}

\begin{table}[!ht]
\tabcolsep 0pt 
\caption{Some resonances formed in $\pi\pi \leftrightarrow \rho\rho$,
$\pi K^\ast \leftrightarrow \rho K$ and $\pi\pi \leftrightarrow  K \bar{K}$.}
\label{rhorho} 
\vspace*{-12pt}
\begin{center}
\def\temptablewidth{0.7\textwidth}
{\rule{\temptablewidth}{1pt}}
\begin{tabular*}{\temptablewidth}{@{\extracolsep{\fill}}llll}
Name & $\Gamma~(\rm{MeV})$ & $B_{\pi\pi}$ & $B_{\rho\rho}$ \\ 
\hline
$f_0(1370) $ & 370 & 0.26 & 0.208\\
$\rho(1450) $ & 147 & 0.0672 & 0.02\\
\hline
Name & $\Gamma~(\rm{MeV})$ & $B_{\pi K^\ast }$ & $B_{\rho K}$ \\ 
\hline
$K_1(1400) $ & 174 & 0.94 & 0.3\\
$K^\ast (1410) $ & 232 & 0.4 & 0.07\\
$K^\ast(1680) $ & 322 & 0.299 & 0.314\\
\hline
Name & $\Gamma~(\rm{MeV})$ & $B_{\pi\pi}$ & $B_{K \bar{K}}$ \\ 
\hline
$f_0(980) $ & 70 & 0.755 & 0.245\\
$f_0(1370) $ &  370 & 0.203 & 0.35\\
$\rho(1450) $ & 147 & 0.0672 & 0.0016\\
$\rho(1700) $ & 250 & 0.2345 & 0.0412\\
$f_2(1810) $ & 197 & 0.0048 & 0.003
\end{tabular*}
{\rule{\temptablewidth}{1pt}}
\end{center}
\end{table}

\begin{table}[htbp]
\centering \caption{$\bar R_{\pi}$, $\bar R_{\rho}$, $\bar R_{K}$ and 
$\bar R_{K^*}$ irrelevant and relevant to the $2 \leftrightarrow 1$ processes 
are listed in the middle and right four columns, respectively. Initial 
fugacities are in the left four columns.}
\label{fug+R_1}
\begin{tabular*}{16cm}{@{\extracolsep{\fill}}cccccccccccc}
  \hline
    $\lambda_\pi$ & $\lambda_\rho$ & $\lambda_K$ & $\lambda_{K^\ast}$
& $\bar {R}_{\pi}$  & $\bar {R}_{\rho}$  & $\bar {R}_K$ & $\bar {R}_{K^*}$  
& $\bar {R}_{\pi}$  & $\bar {R}_{\rho}$  & $\bar {R}_K$ & $\bar {R}_{K^*}$

\\
  \hline
  0.35 & 0.15 & 0.25 & 0.15 & 0.43 & 0.18 & 1.14 & 1.80 &
  1.34 & 1.54 & 0.95 & 1.02
\\
  0.35 & 0.15 & 0.25 & 0.55 & 1.63 & 0.19 & 0.30 & 0.20 &
  0.99 & 2.03 & 4.79 & 0.93
\\
  0.35 & 0.15 & 0.25 & 0.95 & 0.36 & 0.19 & 0.27 & 0.34 &
  0.87 & 2.81 & 1.17 & 0.90
\\
  0.35 & 0.15 & 0.55 & 0.15 & 1.19 & 0.22 & 0.01 & 2.52 &
  1.75 & 6.89 & 0.53 & 1.06
\\
  0.35 & 0.15 & 0.55 & 0.55 & 0.45 & 0.23 & 0.17 & 0.03 &
  1.34 & 2.68 & 0.36 & 0.85
\\
  0.35 & 0.15 & 0.55 & 0.95 & 0.27 & 0.21 & 2.86 & 0.04 &
  0.93 & 2.04 & 6.22 & 0.83
\\
  0.35 & 0.15 & 0.85 & 0.15 & 5.28 & 0.50 & 0.01 & 2.56 &
  4.49 & 2.42 & 0.50 & 1.72
\\
  0.35 & 0.15 & 0.85 & 0.55 & 0.54 & 0.74 & 0.04 & 0.08 &
  2.23 & 2.71 & 0.35 & 0.77
\\
  0.35 & 0.15 & 0.85 & 0.95 & 0.29 & 0.22 & 0.09 & 0.03 &
  2.34 & 4.07 & 0.27 & 0.77
\\
  0.35 & 0.45 & 0.25 & 0.15 & 0.61 & 0.28 & 0.22 & 0.18 &
  1.07 & 1.00 & 0.98 & 1.16
\\
  0.35 & 0.45 & 0.25 & 0.55 & 0.44 & 2.76 & 0.27 & 0.39 &
  1.00 & 5.32 & 1.66 & 0.96
\\
  0.35 & 0.45 & 0.25 & 0.95 & 0.40 & 2.16 & 0.22 & 0.42 &
  0.93 & 1.41 & 1.52 & 0.94
\\
  0.35 & 0.45 & 0.55 & 0.15 & 0.28 & 0.21 & 0.13 & 2.07 &
  1.27 & 0.97 & 0.77 & 1.77
\\
  0.35 & 0.45 & 0.55 & 0.55 & 0.74 & 0.26 & 0.10 & 0.13 &
  1.48 & 1.04 & 0.67 & 0.93
\\
  0.35 & 0.45 & 0.55 & 0.95 & 0.50 & 0.66 & 0.08 & 0.22 &
  0.99 & 1.29 & 0.45 & 0.90
\\
  0.35 & 0.45 & 0.85 & 0.15 & 0.45 & 0.20 & 0.10 & 5.44 &
  1.98 & 0.97 & 0.68 & 0.80
\\
  0.35 & 0.45 & 0.85 & 0.55 & 0.26 & 0.21 & 0.08 & 0.10 &
  1.98 & 1.01 & 0.60 & 0.87
\\
  0.35 & 0.45 & 0.85 & 0.95 & 0.99 & 0.29 & 0.07 & 0.11 &
  1.75 & 1.13 & 0.50 & 0.85
\\
  0.35 & 0.75 & 0.25 & 0.15 & 7.50 & 0.28 & 0.25 & 0.40 &
  1.04 & 0.99 & 1.00 & 1.25
\\
  0.35 & 0.75 & 0.25 & 0.55 & 0.42 & 0.41 & 0.34 & 0.42 &
  1.00 & 1.02 & 1.38 & 0.97
\\
  0.35 & 0.75 & 0.25 & 0.95 & 0.42 & 0.57 & 0.58 & 0.43 &
  0.95 & 1.07 & 1.31 & 0.95
\\
  0.35 & 0.75 & 0.55 & 0.15 & 0.20 & 0.24 & 0.18 & 0.47 &
  1.16 & 0.96 & 0.83 & 1.72
\\
  0.35 & 0.75 & 0.55 & 0.55 & 0.92 & 0.28 & 0.19 & 0.22 &
  1.12 & 0.98 & 0.78 & 0.96
\\
  0.35 & 0.75 & 0.55 & 0.95 & 1.32 & 0.34 & 0.19 & 0.30 &
  1.03 & 1.01 & 0.70 & 0.93
\\
  0.35 & 0.75 & 0.85 & 0.15 & 0.32 & 0.22 & 0.15 & 0.69 &
  1.45 & 0.94 & 0.74 & 1.11
\\
  0.35 & 0.75 & 0.85 & 0.55 & 0.39 & 0.25 & 0.15 & 0.56 &
  1.42 & 0.96 & 0.70 & 1.11
\\
  0.35 & 0.75 & 0.85 & 0.95 & 4.53 & 0.28 & 0.16 & 0.19 &
  1.72 & 0.98 & 0.64 & 0.89
\\
  \hline
\end{tabular*}
\end{table}

\begin{table}[htbp]
\centering \caption{The same as Table 6.}
\label{fug+R_2}
\begin{tabular*}{16cm}{@{\extracolsep{\fill}}cccccccccccc}
  \hline
    $\lambda_\pi$ & $\lambda_\rho$ & $\lambda_K$ & $\lambda_{K^\ast}$
& $\bar {R}_{\pi}$  & $\bar {R}_{\rho}$  & $\bar {R}_K$ & $\bar {R}_{K^*}$  
& $\bar {R}_{\pi}$  & $\bar {R}_{\rho}$  & $\bar {R}_K$ & $\bar {R}_{K^*}$

\\
  \hline
  0.65 & 0.15 & 0.25 & 0.15 & 0.15 & 0.17 & 0.14 & 1.74 &
  0.68 & 1.44 & 0.70 & 1.08
\\
  0.65 & 0.15 & 0.25 & 0.55 & 0.17 & 0.20 & 0.20 & 1.61 &
  1.25 & 0.73 & 0.89 & 0.87
\\
  0.65 & 0.15 & 0.25 & 0.95 & 0.11 & 0.20 & 0.22 & 1.36 &
  0.98 & 0.74 & 0.94 & 0.83
\\
  0.65 & 0.15 & 0.55 & 0.15 & 10.48 & 0.16 & 0.14 & 5.73 &
  0.92 & 2.30 & 7.31 & 1.11
\\
  0.65 & 0.15 & 0.55 & 0.55 & 0.51 & 0.21 & 6.34 & 0.07 &
  0.86 & 0.68 & 5.43 & 1.00
\\
  0.65 & 0.15 & 0.55 & 0.95 & 2.90 & 0.21 & 0.51 & 0.08 &
  0.87 & 0.72 & 5.80 & 0.76
\\
  0.65 & 0.15 & 0.85 & 0.15 & 1.50 & 0.18 & 0.03 & 4.62 &
  1.34 & 1.42 & 0.34 & 1.74
\\
  0.65 & 0.15 & 0.85 & 0.55 & 2.15 & 0.21 & 0.24 & 0.11 &
  0.78 & 1.09 & 0.41 & 1.35
\\
  0.65 & 0.15 & 0.85 & 0.95 & 1.71 & 0.22 & 3.52 & 0.09 &
  0.83 & 0.69 & 12.40 & 0.69
\\
  0.65 & 0.45 & 0.25 & 0.15 & 0.14 & 2.44 & 2.00 & 0.09 &
  1.73 & 1.67 & 3.07 & 1.43
\\
  0.65 & 0.45 & 0.25 & 0.55 & 0.59 & 0.13 & 0.18 & 0.69 &
  1.37 & 1.64 & 0.74 & 0.96
\\
  0.65 & 0.45 & 0.25 & 0.95 & 1.03 & 0.15 & 0.21 & 0.62 &
  0.98 & 1.99 & 0.86 & 0.93
\\
  0.65 & 0.45 & 0.55 & 0.15 & 0.36 & 0.28 & 0.13 & 0.33 &
  1.62 & 1.04 & 0.72 & 0.92
\\
  0.65 & 0.45 & 0.55 & 0.55 & 2.09 & 1.14 & 0.80 & 0.13 &
  1.52 & 1.60 & 1.08 & 0.90
\\
  0.65 & 0.45 & 0.55 & 0.95 & 0.44 & 0.17 & 3.52 & 0.22 &
  0.97 & 2.56 & 2.85 & 0.87
\\
  0.65 & 0.45 & 0.85 & 0.15 & 7.58 & 0.23 & 0.09 & 0.51 &
  1.62 & 0.97 & 0.60 & 1.19
\\
  0.65 & 0.45 & 0.85 & 0.55 & 0.37 & 4.12 & 0.03 & 0.47 &
  4.25 & 1.53 & 0.48 & 1.10
\\
  0.65 & 0.45 & 0.85 & 0.95 & 0.29 & 0.30 & 0.08 & 0.07 &
  1.68 & 1.91 & 0.35 & 0.80
\\
  0.65 & 0.75 & 0.25 & 0.15 & 1.85 & 0.52 & 2.24 & 0.12 &
  1.18 & 1.05 & 3.80 & 42.57
\\
  0.65 & 0.75 & 0.25 & 0.55 & 0.76 & 3.09 & 0.10 & 0.60 &
  1.06 & 1.46 & 1.99 & 0.98
\\
  0.65 & 0.75 & 0.25 & 0.95 & 0.54 & 4.69 & 0.19 & 0.53 &
  0.98 & 1.85 & 0.89 & 0.95
\\
  0.65 & 0.75 & 0.55 & 0.15 & 0.27 & 0.27 & 0.20 & 0.30 &
  1.27 & 0.97 & 0.84 & 1.13
\\
  0.65 & 0.75 & 0.55 & 0.55 & 0.93 & 0.57 & 0.16 & 0.46 &
  1.15 & 1.09 & 0.78 & 1.01
\\
  0.65 & 0.75 & 0.55 & 0.95 & 0.44 & 1.58 & 0.51 & 0.33 &
  0.98 & 1.90 & 0.60 & 0.92
\\
  0.65 & 0.75 & 0.85 & 0.15 & 0.37 & 0.24 & 0.15 & 0.42 &
  1.65 & 0.94 & 0.71 & 0.85
\\
  0.65 & 0.75 & 0.85 & 0.55 & 0.26 & 0.26 & 0.13 & 0.82 &
  1.51 & 0.99 & 0.65 & 1.71
\\
  0.65 & 0.75 & 0.85 & 0.95 & 0.88 & 0.45 & 0.10 & 0.17 &
  1.19 & 1.16 & 0.54 & 0.87
\\
  \hline
\end{tabular*}
\end{table}

\begin{table}[htbp]
\centering \caption{The same as Table 6.}
\label{fug+R_3}
\begin{tabular*}{16cm}{@{\extracolsep{\fill}}cccccccccccc}
  \hline
    $\lambda_\pi$ & $\lambda_\rho$ & $\lambda_K$ & $\lambda_{K^\ast}$
& $\bar {R}_{\pi}$  & $\bar {R}_{\rho}$  & $\bar {R}_K$ & $\bar {R}_{K^*}$  
& $\bar {R}_{\pi}$  & $\bar {R}_{\rho}$  & $\bar {R}_K$ & $\bar {R}_{K^*}$

\\
  \hline
  0.95 & 0.15 & 0.25 & 0.15 & 0.12 & 0.17 & 0.09 & 3.27 &
  0.86 & 0.88 & 0.65 & 1.15
\\
  0.95 & 0.15 & 0.25 & 0.55 & 0.16 & 0.20 & 0.17 & 2.74 &
  0.85 & 0.84 & 0.78 & 1.44
\\
  0.95 & 0.15 & 0.25 & 0.95 & 0.17 & 0.20 & 0.20 & 1.32 &
  1.26 & 0.83 & 0.85 & 1.30
\\
  0.95 & 0.15 & 0.55 & 0.15 & 0.23 & 0.16 & 3.95 & 0.34 &
  0.92 & 0.90 & 4.40 & 0.96
\\
  0.95 & 0.15 & 0.55 & 0.55 & 0.26 & 0.21 & 0.34 & 0.08 &
  0.95 & 0.86 & 1.38 & 1.64
\\
  0.95 & 0.15 & 0.55 & 0.95 & 0.28 & 0.21 & 0.31 & 0.21 &
  1.75 & 0.84 & 1.29 & 1.88
\\
  0.95 & 0.15 & 0.85 & 0.15 & 4.44 & 0.14 & 0.11 & 0.95 &
  0.95 & 0.90 & 1.54 & 1.03
\\
  0.95 & 0.15 & 0.85 & 0.55 & 0.67 & 0.21 & 3.62 & 0.10 &
  1.02 & 0.86 & 8.08 & 1.67
\\
  0.95 & 0.15 & 0.85 & 0.95 & 0.73 & 0.22 & 8.17 & 0.11 &
  1.98 & 0.84 & 8.36 & 2.43
\\
  0.95 & 0.45 & 0.25 & 0.15 & 0.06 & 0.08 & 0.03 & 0.03 &
  2.55 & 1.32 & 0.30 & 1.26
\\
  0.95 & 0.45 & 0.25 & 0.55 & 0.06 & 0.17 & 0.16 & 33.07 &
  1.73 & 1.41 & 0.65 & 0.94
\\
  0.95 & 0.45 & 0.25 & 0.95 & 0.04 & 0.18 & 0.19 & 1.84 &
  1.31 & 2.17 & 0.77 & 0.86
\\
  0.95 & 0.45 & 0.55 & 0.15 & 0.14 & 1.42 & 0.45 & 0.24 &
  1.74 & 1.25 & 1.15 & 0.89
\\
  0.95 & 0.45 & 0.55 & 0.55 & 0.15 & 0.19 & 0.54 & 0.34 &
  1.13 & 1.21 & 2.25 & 1.03
\\
  0.95 & 0.45 & 0.55 & 0.95 & 0.10 & 0.21 & 0.33 & 0.28 &
  1.05 & 1.13 & 3.67 & 0.81
\\
  0.95 & 0.45 & 0.85 & 0.15 & 0.97 & 0.75 & 0.08 & 0.37 &
  1.15 & 1.23 & 0.47 & 0.96
\\
  0.95 & 0.45 & 0.85 & 0.55 & 0.99 & 0.31 & 0.19 & 1.44 &
  1.02 & 1.47 & 0.57 & 1.22
\\
  0.95 & 0.45 & 0.85 & 0.95 & 2.48 & 0.24 & 2.60 & 0.05 &
  0.82 & 1.21 & 15.76 & 1.01
\\
  0.95 & 0.75 & 0.25 & 0.15 & 0.17 & 5.86 & 0.20 & 0.08 &
  2.13 & 2.59 & 4.36 & 1.46
\\
  0.95 & 0.75 & 0.25 & 0.55 & 0.46 & 0.08 & 0.14 & 1.11 &
  1.76 & 2.07 & 0.41 & 0.99
\\
  0.95 & 0.75 & 0.25 & 0.95 & 15.83 & 0.12 & 0.19 & 0.72 &
  1.49 & 2.32 & 0.68 & 0.93
\\
  0.95 & 0.75 & 0.55 & 0.15 & 0.96 & 0.41 & 0.29 & 0.25 &
  1.90 & 1.09 & 0.93 & 1.12
\\
  0.95 & 0.75 & 0.55 & 0.55 & 4.09 & 4.11 & 0.28 & 1.85 &
  1.51 & 1.40 & 0.85 & 1.16
\\
  0.95 & 0.75 & 0.55 & 0.95 & 8.19 & 0.14 & 1.61 & 0.41 &
  1.13 & 1.58 & 1.39 & 0.88
\\
  0.95 & 0.75 & 0.85 & 0.15 & 0.32 & 0.28 & 0.16 & 0.34 &
  1.57 & 0.88 & 0.66 & 0.88
\\
  0.95 & 0.75 & 0.85 & 0.55 & 0.25 & 0.78 & 0.09 & 1.51 &
  1.31 & 1.16 & 0.54 & 0.92
\\
  0.95 & 0.75 & 0.85 & 0.95 & 0.70 & 0.47 & 0.84 & 0.18 &
  1.14 & 2.50 & 1.15 & 0.78
\\
  \hline
\end{tabular*}
\end{table}

\newpage
\begin{figure}
\centering
\includegraphics[scale=1.0,width=15cm]{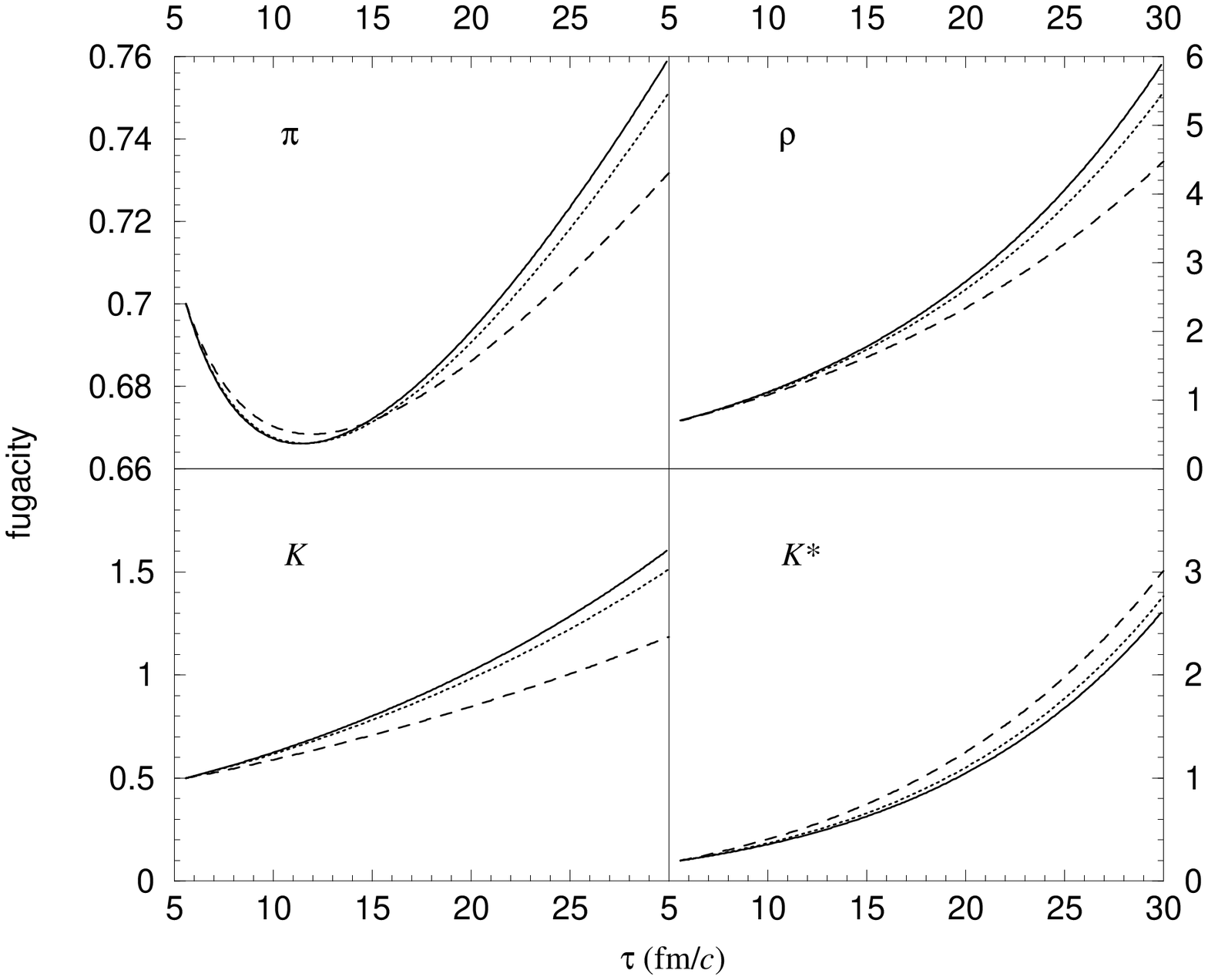}
\caption{Time dependence of $\lambda_{\pi}$, $\lambda_{\rho}$, $\lambda_{K}$
and $\lambda_{K^{\ast}}$ without the source terms (solid curves), with the
quark-interchange, annihilation and resonant 
processes (dashed curves) and with only the quark-interchange processes (dotted
curves).
}\label{fig1}
\end{figure}


\newpage
\begin{thebibliography}{99}
\addtolength{\itemsep}{-0.6 em}
\bibitem{RQMD1}H. Sorge, H. St\"{o}cker, W. Greiner, Ann. Phys. 
\textbf{192}, 266 (1989).
\bibitem{RQMD2}H. Sorge, Phys. Rev. C \textbf{52}, 3291 (1995).
\bibitem{ART1}B.-A. Li, C.M. Ko, Phys. Rev. C \textbf{52}, 2037 (1995).
\bibitem{ART2}Z.-W. Lin, C.M. Ko, B.-A. Li, B. Zhang, S. Pal, Phys. Rev. C
\textbf{72}, 064901 (2005).
\bibitem{HSD}W. Cassing, E.L. Bratkovskaya, Phys. Rep. \textbf{308}, 65 
(1999).
\bibitem{UrQMD1}S.A. Bass {\it et al.}, Prog. Part. Nucl. Phys. \textbf{41}, 
255 (1998). 
\bibitem{UrQMD2}C. Nonaka, S.A. Bass, Phys. Rev. C \textbf{75}, 014902 (2007).
\bibitem{ARC1}D.E. Kahana, S.H. Kahana, Phys. Rev. C \textbf{58}, 3574 (1998). 
\bibitem{ARC2}D.E. Kahana, S.H. Kahana, Phys. Rev. C \textbf{59}, 1651 (1999).
\bibitem{PACIAE1}B.-H. Sa, A. Tai, Comput. Phys. Commun. \textbf{90}, 121
(1995).
\bibitem{PACIAE2}A. Tai, B.-H. Sa, Comput. Phys. Commun. \textbf{116}, 353 
(1999).
\bibitem{Humanic}T.J. Humanic, Phys. Rev. C \textbf{57}, 866 (1998).
\bibitem{Nara1}Y. Nara, N. Otuka, A. Ohnishi, K. Niita, S. Chiba, Phys. Rev.
C \textbf{61}, 024901 (1999).
\bibitem{Nara2}T. Hirano, U. Heinz, D. Kharzeev, R. Lacey,
Y. Nara, Phys. Lett. B \textbf{636}, 299 (2006).
\bibitem{Rus02}L. Alvarez-Ruso, V. Koch, Phys. Rev. C \textbf{65}, 054901 
(2002).
\bibitem{Gav91}S. Gavin, Nucl. Phys. B \textbf{351}, 561 (1991).
\bibitem{Son97}C. Song, V. Koch, Phys. Rev. C \textbf{55}, 3026 (1997).
\bibitem{Pra99}S. Pratt, K. Haglin, Phys. Rev. C \textbf{59}, 3304 (1999).
\bibitem{SSAdl04}S.S. Adler {\it et al.} (PHENIX Collaboration), Phys. Rev. 
C \textbf{69}, 034909 (2004).
\bibitem{CAdl04}C. Adler {\it et al.} (STAR Collaboration), Phys. Lett. B
\textbf{595}, 143 (2004).
\bibitem{Ars05}I. Arsene {\it et al.} (BRAHMS Collaboration), Nucl. Phys. A
\textbf{757}, 1 (2005).
\bibitem{Bac05}B. B. Back {\it et al.} (PHOBOS Collaboration), Nucl. Phys. A
\textbf{757}, 28 (2005).
\bibitem{Ada04}J. Adams {\it et al.} (STAR Collaboration), Phys. Rev. Lett. 
\textbf{92},  092301 (2004).
\bibitem{AT}C. Amsler, N.A. T\"{o}rnqvist, Phys. Rep. \textbf{389}, 61 
(2004).
\bibitem{Bugg}D.V. Bugg, Phys. Rep. \textbf{397}, 257 (2004).
\bibitem{Li07}Y.-Q. Li, X.-M. Xu, Nucl. Phys. A \textbf{794}, 210 (2007).
\bibitem{Bar92A}T. Barnes, E.S. Swanson, Phys. Rev. D \textbf{46}, 131 (1992). 
\bibitem{Bar92B}E.S. Swanson, Ann. Phys. \textbf{220}, 73 (1992).
\bibitem{Bjo83}J.D. Bjorken, Phys. Rev. D \textbf{27}, 140 (1983).
\bibitem{Sor95}H. Sorge, Phys. Lett. B \textbf{402}, 251 (1997).
\bibitem{Bra98}L.V. Bravina {\it et al.}, Phys. Rev. C \textbf{60}, 024904 
(1999).
\bibitem{Kolb1}P.F. Kolb, J. Sollfrank, U. Heinz, Phys. Lett. B \textbf{459},
 667 (1999).
\bibitem{Kolb2}P.F. Kolb, J. Sollfrank, U. Heinz, Phys. Rev. C \textbf{62}, 
054909 (2000).
\bibitem{Buc81}W. Buchm\"{u}ller, S.-H.H. Tye, Phys. Rev. D \textbf{24}, 
132 (1981).
\bibitem{Xu02}X.-M. Xu, Nucl. Phys. A \textbf{697}, 825 (2002).
\bibitem{Bar03}T. Barnes, E.S. Swanson, C.-Y. Wong, X.-M. Xu, Phys. Rev. C 
\textbf{68}, 014903 (2003).
\bibitem{Bro91}G.E. Brown, C.M. Ko, Z.G. Wu, L.H. Xia, Phys. Rev. C 
\textbf{43}, 1881 (1991).
\bibitem{Cas97}W. Cassing, E.L. Bratkovskaya, U. Mosel, S. Teis,
 A. Sibirtsev, Nucl. Phys. A \textbf{614}, 415 (1997).
\bibitem{PDG}Particle Data Group, W.-M. Yao {\it et al.}, J. Phys. G 
\textbf{33}, 1 (2006).
\bibitem{Kar01}F. Karsch, E. Laermann, A. Peikert, Nucl. Phys. B {\bf 605}, 
579 (2001).
\bibitem{HIJING1}X.-N. Wang, M. Gyulassy, Phys. Rev. D {\bf 44}, 3501 (1991).
\bibitem{HIJING2}M. Gyulassy, X.-N. Wang, Comput. Phys. Commun. {\bf 83}, 307 
(1994).
\bibitem{HIJING3}X.-N. Wang, Phys. Rep. {\bf 280}, 287 (1997).
\bibitem{Xu96}X.-M. Xu, D. Kharzeev, H. Satz, X.-N. Wang, Phys. Rev. C 
\textbf{53}, 3051 (1996).
\bibitem{Biro99}T.S. Bir$\rm \acute o$, P. L$\rm \acute e$vai, 
J. Zim$\rm \acute a$nyi, Phys. Rev. C \textbf{59}, 1574 (1999).
\bibitem{Baran}O. Barannikova, for the STAR Collaboration, 
arXiv:nucl-ex/0403014v1.
\bibitem{Lev95}P. L$\rm \acute e$vai, B. M\"{u}ller, X.-N. Wang, Phys. Rev. C 
\textbf{51}, 3326 (1995).
\bibitem{Fla84}V. Flaminio, W.G. Moorhead, D.R.O. Morrison, N. Rivoire, CERN,
Geneva Report No. CERN-HERA-84-01, 1984. 
\bibitem{Ger86}H. von Gersdorff, L. McLerran, M. Kataja, P.V. Ruuskanen, Phys. 
Rev. D \textbf{34}, 794 (1986).
\end{thebibliography}
\end{document}